%% file: ms.tex
\documentclass[10pt,journal,compsoc]{IEEEtran}

\ifCLASSOPTIONcompsoc
  \usepackage[nocompress]{cite}
\else
  \usepackage{cite}
\fi

\newcommand\MYhyperrefoptions{bookmarks=true,bookmarksnumbered=true,
pdfpagemode={UseOutlines},plainpages=false,pdfpagelabels=true,
colorlinks=true,linkcolor={black},citecolor={black},urlcolor={black},
pdftitle={FAME: 3D Shape Generation via Functionality-Aware Model Evolution},
pdfsubject={Geometric Modeling},
pdfauthor={Yanran Guan, Han Liu, Kun Liu, Kangxue Yin, Ruizhen Hu, Oliver van Kaick, Yan Zhang, Ersin Yumer, Nathan Carr, Radomir Mech, Hao Zhang},
pdfkeywords={cross-category hybrids, functionality-aware shape modeling, functionality partial matching, set evolution}}

\usepackage{booktabs} 

\usepackage[ruled,vlined]{algorithm2e} 

\SetAlFnt{\small}
\SetAlCapFnt{\small}
\SetAlCapNameFnt{\small}
\SetAlCapHSkip{0pt}
\IncMargin{-\parindent}

\usepackage{amsmath, amsthm, amsfonts, amssymb}
\usepackage{mathrsfs}
\usepackage{graphicx}
\usepackage{textcomp}
\usepackage{wrapfig}
\usepackage{subfig}
\usepackage{parskip}
\usepackage{color}
\usepackage{dsfont}
\usepackage{siunitx}
\usepackage[\MYhyperrefoptions]{hyperref}

\definecolor{black}{rgb}{0,0,0}
\definecolor{turquoise}{cmyk}{0.65,0,0.1,0.1}
\definecolor{purple}{rgb}{0.65,0,0.65}
\definecolor{dark_green}{rgb}{0, 0.5, 0}
\definecolor{orange}{rgb}{0.8, 0.6, 0.2}
\definecolor{red}{rgb}{0.8, 0.2, 0.2}
\definecolor{brown}{rgb}{0.5, 0.16, 0.16}

\newcommand{\mypara}[1]{\textit{#1}.}

\begin{document}

\title{FAME: 3D Shape Generation via \\ Functionality-Aware Model Evolution}

\author{Yanran Guan,
Han Liu,
Kun Liu,
Kangxue Yin,
Ruizhen Hu,
Oliver van Kaick, \\
Yan Zhang,
Ersin Yumer,
Nathan Carr,
Radomir Mech,
Hao Zhang
\IEEEcompsocitemizethanks{
    \IEEEcompsocthanksitem Yanran Guan and Oliver van Kaick are with the School of Computer Science, Carleton University, Canada.
    \IEEEcompsocthanksitem Han Liu is with Electronic Arts, USA. This work was carried out when she was a postdoc at Simon Fraser University and Carleton University.
    \IEEEcompsocthanksitem Kun Liu and Yan Zhang are with the Department of Computer Science and Technology, Nanjing University, China.
    \IEEEcompsocthanksitem Kangxue Yin is with NVIDIA, Canada.
    \IEEEcompsocthanksitem Ruizhen Hu is with the College of Computer Science and Software Engineering, Shenzhen University, China.
    \IEEEcompsocthanksitem Ersin Yumer is with Uber ATG, USA.
    \IEEEcompsocthanksitem Nathan Carr and Radomir Mech are with Adobe, USA.
    \IEEEcompsocthanksitem Hao Zhang is with the School of Computing Science, Simon Fraser University, Canada.
}
\thanks{Corresponding author: Ruizhen Hu (ruizhen.hu@gmail.com)}}


\IEEEtitleabstractindextext{%
\begin{abstract}
\input{abstract}
\end{abstract}

}

\maketitle

\IEEEdisplaynontitleabstractindextext

\IEEEpeerreviewmaketitle


\input{intro}

\input{related}

\input{evolution}

\input{function}
\input{results}
\input{future}
\input{ack}

\ifCLASSOPTIONcaptionsoff
  \newpage
\fi

\bibliographystyle{abbrv-doi-narrow}
\bibliography{references}


\end{document}

%% file: abstract.tex
We introduce a modeling tool which can evolve a set of 3D objects in a \emph{functionality-aware} 
manner. Our goal is for the evolution to generate large and diverse sets of plausible 3D 
objects for data augmentation, constrained modeling, as well as open-ended exploration 
to possibly inspire new designs.
Starting with an initial population of 3D objects belonging to \emph{one or more} functional
categories, we evolve the shapes through part recombination to produce
generations of \emph{hybrids} or crossbreeds between parents from the heterogeneous 
shape collection.
Evolutionary selection of offsprings is guided both by a \emph{functional plausibility score} 
derived from functionality analysis of shapes in the initial population and user preference,
as in a design gallery.
Since cross-category hybridization may result in offsprings not belonging to any of the known functional 
categories, we develop a means for \emph{functionality partial matching} to evaluate functional plausibility
on partial shapes.
We show a variety of plausible hybrid shapes generated by our functionality-aware model evolution, which
can complement existing datasets as training data and boost the performance of contemporary data-driven 
segmentation schemes, especially in challenging cases.
Our tool supports constrained modeling, allowing users to restrict or steer the model evolution with 
functionality labels. At the same time, unexpected yet functional object prototypes can emerge 
during open-ended exploration owing to structure breaking when evolving a heterogeneous collection.
%
%
%


%% file: intro.tex

\section{Introduction}
\label{sec:intro}

\IEEEPARstart{3}{D} object modeling lies at the core of computer graphics.  With the resurgence of VR/AR, deep learning,
and 3D printing technologies in recent years, there is a strong push to develop innovative tools for 
3D content creation, which can play key roles in data augmentation and the 
design and rapid prototyping of real 3D products.
Current modeling tools have mainly focused on appearance, style, and aesthetic aspects of the generated
shapes. One important criterion that is largely missing is \emph{functionality}. 
When customizing an existing design or evolving current designs into a new prototype, the most 
basic requirement is for the final products to serve their intended functional purposes.
In addition, rapid developments in geometric deep learning are placing an ever increasing demand 
for 3D models. A tool that is capable of producing functionally plausible 3D shapes in large volumes and varieties is highly desirable.

\input{figures/teaser}

\input{figures/overview}

In this paper, we introduce a modeling tool which can \emph{evolve} a set of 3D objects in a 
\emph{functionality-aware} manner. We call our tool ``FAME'', short for \emph{functionality-aware model evolution}. Given an initial population of 3D objects belonging to 
\emph{one or more} functional categories, FAME produces generations and generations of 
\emph{functionally plausible hybrids} or crossbreeds between parents from preceding generations.
%
%
For example, we could crossbreed a rocking crib and a chair into a 
combo which can comfort both a parent and a baby; see Fig.~\ref{fig:teaser} (top).
%
%
Taking an evolutionary approach to shape generation emphasizes that modeling is a continuous 
process of \emph{controlled diversification}. 
Our main goal is for the evolutionary modeling tool to produce many plausible 3D 
prototypes that are both ``\emph{fit}'' and ``\emph{diverse}''~\cite{xu2012}, so that they can
augment existing datasets to boost learning algorithms and allow both constrained modeling and 
open-ended exploration to possibly inspire new designs.

The main challenges to functionality-aware model evolution are two-fold. The first is attributed to
\emph{cross-category} modeling. Unlike all previous works on co-analysis and data-driven shape 
processing~\cite{xu2016}, our tool works with a \emph{heterogeneous} shape collection and
composes parts from different object or functional categories. As a result, the
well-known notion of structure-preserving shape modeling~\cite{mitra2013} can no longer be
strictly adhered to; some levels of \emph{structure breaking} must occur.
The second challenge is to adapt and enhance functionality models which were designed for discriminative
analysis~\cite{hu2015,hu2016,pirk2016} to serve shape modeling. In particular, we must address the issue
with new object categories arising from cross-category hybrids whose functionality models could not be 
learned in advance.

Starting with an initial population of segmented 3D objects, 
we first obtain a functional understanding of the input. 
Then we evolve the population, where at each iteration, the objects undergo \emph{stochastic} part recombination, 
mimicking crossovers in evolutionary biology. Part deformation is applied to properly scale and connect the parts after a crossover. 
Evolutionary selection of offsprings, i.e., the fitness criterion, is guided both by a \emph{functional plausibility score} 
derived from functionality analysis of the initial population and user preference, as in a design 
gallery~\cite{marks1997,xu2012}. Specifically, users can express preferences over liked offsprings,
which will subsequently enter the population as parents while the remaining shapes would go extinct.
In addition, our tool supports constrained modeling, allowing users to restrict or steer the model evolution
with constraints defined via intuitive functionality labels, such as \emph{sitting}, \emph{leaning}, and \emph{storage}.
For example, the user can pick two functionalities and constrain the model evolution to produce offsprings 
possessing the two specified functionalities.


Our evolution operates at the \emph{part group} level, where a part group consists
of one or more related shape parts. Evolution is applied to a set (current population) of shapes, where
each modeling operation is a crossover (i.e., an exchange of part groups between two 3D shapes), or mutation 
(i.e., part insertion into a 3D shape), followed by part (group) deformation and connection to improve plausibility of 
the offspring. At each iteration, we generate models that satisfy the modeling constraints, rank them by their functional 
plausibility scores, and select the top $k$ offsprings while respecting user-expressed preferences to enter
the evolving population. Since cross-category hybridization may result in offsprings not belonging to any of 
the known functional categories, we develop a means for \emph{functionality partial matching} by \emph{localizing} 
the functionality models developed by Hu et al.~\cite{hu2016} from the category/object level down to 
patch/part level. We then combine the functionality partial matching scores against the known functional categories 
into a plausibility score for the offspring;
see Fig.~\ref{fig:overview}. 



%
%

Our main contributions are summarized and reasoned, in the context of the state-of-the-art, as follows:
\begin{itemize}
\item We present the first 3D shape modeling tool based on functionality-aware model evolution (FAME).
By evolving sets of 3D shapes progressively, FAME enables the generation of \emph{large volumes} 
of \emph{functionally plausible} and \emph{diverse} hybrid objects for the first time, leading to
applications such as data augmentation, constrained modeling, and open-ended design exploration.
\item We generate new shapes via part recombination, possibly across object caragories. 
Importantly, our method goes beyond exchanging parts sharing the same functionality label, since merely
having the functional parts in a shape does \emph{not} imply that the resulting object would fulfill its
intended function, as exemplified by the well-known ``impossible teapot'' design~\cite{norman02}. To this end,
we develop a novel means of evaluating the functionality of the assembled models in a global manner to enforce a proper configuration of the functional parts.
\item To enable functionality evaluation of new hybirds which may not belong to any known
object categories, we introduce a key new concept, namely \emph{functionality partial matching}, which 
analyzes functional plausibility of new shapes not as whole, but in parts, with respect to learned 
functionality models.
\item Our interactive modeling tool requires only \emph{light} user interactions in the
design gallery. It supports both open-ended object exploration and constrained modeling
based on user-prescribed functionality labels for the target shapes, e.g., \emph{sitting} and \emph{rolling}.
\end{itemize}

Typically, the initial population only consists of a handful of 3D objects from one or more functional categories. The shapes need to be provided with an input segmentation and consistent alignment. However, there are no strong requirements on the quality of the triangle meshes.
The continuous evolution is able to produce a large and diverse set of outcomes. 
We show a variety of plausible hybrids generated by our functionality-aware evolution, demonstrating 
controlled design exploration via constrained modeling, as well as the emergence of structure breaking 
and unexpected yet functional object prototypes; see Fig.~\ref{fig:teaser}.

An important utility offered by our new modeling tool is data augmentation. We show that shapes generated
by our tool can complement existing datasets such as ShapeNet~\cite{chang2015_shapeNet} and improve the 
diversity of training shapes and generalizability of data-driven segmentation schemes, 
especially for \emph{atypical} inputs such as shapes with missing parts.



%% file: figures/teaser.tex

\begin{figure}[!t]
  \centering
  \includegraphics[width=\linewidth]{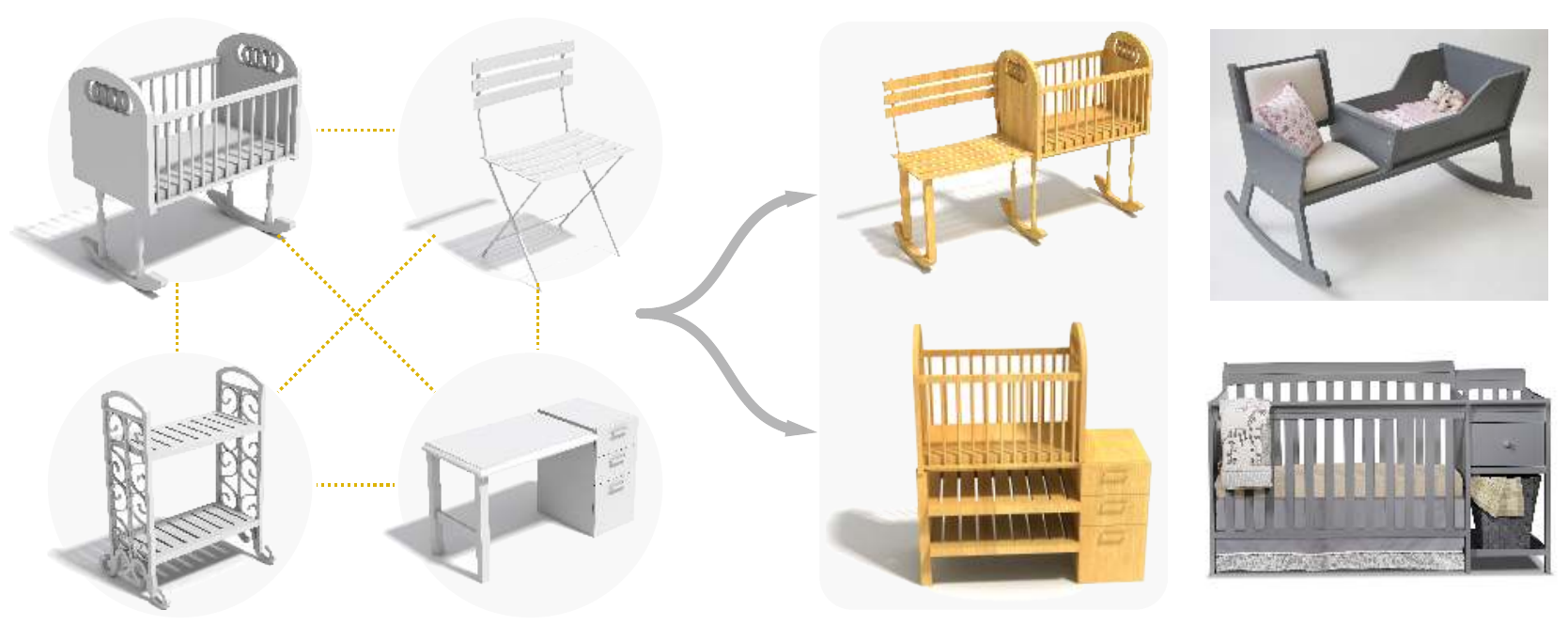}
  \caption{Starting from a \emph{heterogeneous} collection of four objects (left) as initial population, our \emph{functionality-aware 
   evolution} tool can generate a variety of plausible hybrid shapes. Two of them (middle) exhibit strong functional similarity 
   to exiting professional designs (right).}
  \label{fig:teaser}
\end{figure}

%% file: figures/overview.tex
\begin{figure*}[t!]
  \centering
  \includegraphics[width=0.96\linewidth]{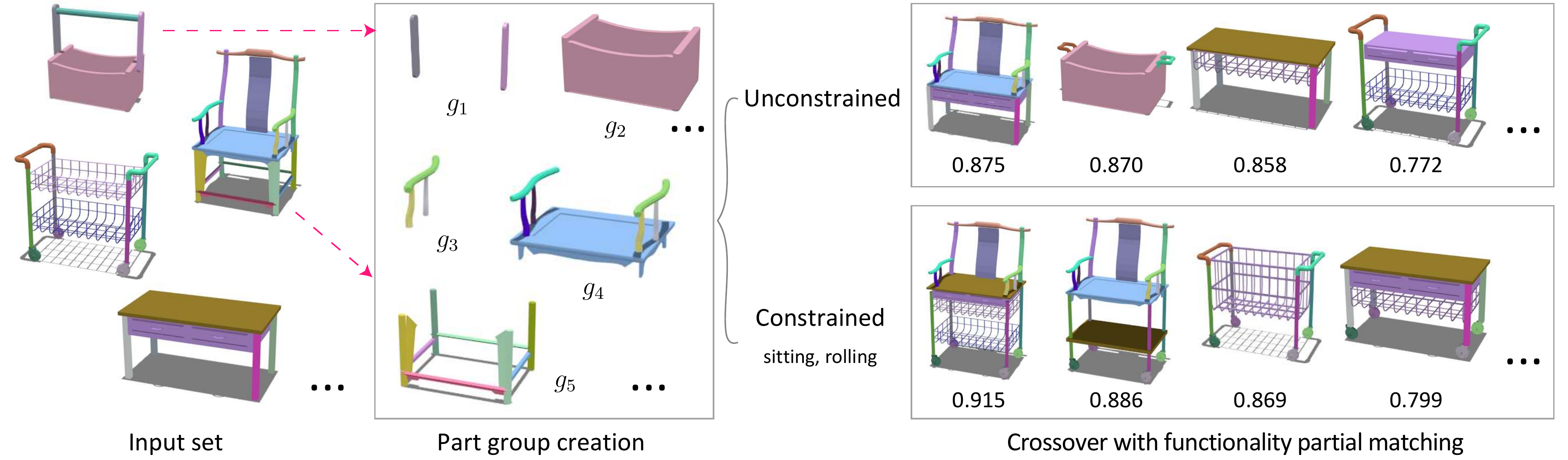}
  \caption{Starting from a set of segmented 3D shapes,
we construct part groups for each shape, where a part group $g_i$ is composed of a combination of one or more shape parts (left).
We apply crossover operations that replace a part group in one shape by a part group of another shape, to create a variety of novel shapes (right). The evolutionary process can be performed in an unconstrained or constrained manner. In the constrained evolution, the user prescribes functionalities that should appear in the output shapes, e.g., \emph{sitting} and \emph{rolling}. The resulting shapes are then ranked according to their functional plausibility. Finally, the user can select shapes to be part of the next generation for further evolution.
}
  \label{fig:overview}
\end{figure*}

%% file: related.tex

\section{Related Work}
\label{sec:related}

Classical shape modeling in computer graphics is subjected to precise geometric constraints or 
controls, while fulfilling low-level modeling criteria such as surface smoothness and detail 
preservation~\cite{sorkine2005}. Recently, much effort has been devoted to structure-preserving
shape processing~\cite{mitra2013}, with more emphasis on analyzing and manipulating
part structures that are characteristic of the class of processed 3D objects. With geometric
modeling playing an increasing role in the design and customization of 3D products, the research 
and development focuses are naturally shifting to higher-level modeling criteria such as 
functionality. 

\mypara{Data-driven model assembly}
In their earlier work on ``modeling by example'', Funkhouser et al.~\cite{funkhouser2004}
pioneered the direction of data-driven 3D object modeling via search-and-assemble schemes
applied to object parts~\cite{xu2016}. Parts from 3D object exemplars can be substituted (into other 
objects) or recombined to form new object prototypes~\cite{shin2007,kraevoy2007,chaudhuri2010,chaudhuri2011,kalogerakis2012,xu2012}.
Mechanisms developed for determining suitable part suggestions include geometric part
similarity~\cite{funkhouser2004,kraevoy2007}, fuzzy correspondence~\cite{kim2012,xu2012}, and more
often probabilistic reasoning learned from examples~\cite{chaudhuri2011,kalogerakis2012}.

Our modeling tool is also data-driven and assembly-based, operating at the part group level. The key 
distinction and novelty of our approach is the incorporation of functionality analysis into the 
process. Moreover, all of the methods mentioned above perform part compositions between shapes
belonging to the same category, while our crossover operations can span different categories.


\mypara{Evolutionary modeling and design}
The seminal works of Karl Sims~\cite{sims1991_sig,sims1994_sig} introduced genetic algorithms 
to the graphics community for the synthesis of novel creatures with desired physical behavior. 
Many follow-up works have appeared since, e.g., the creature academy of Pilat and Jacob~\cite{pilat2008} 
and
automatic design and manufacturing of robotic forms~\cite{lipson2000automatic}. 
%
Most closely related, and inspiring, to our work is the ``fit and diverse'' modeling tool of Xu 
et al.~\cite{xu2012}, which also evolves a set of 3D shapes, 
via part mutation and crossover, while utilizing a design gallery~\cite{marks1997} interface. Again, our 
work distinguishes itself from these previous works by considering object functionality. 
``fit and diverse'' only evolves models of a same category, while our evolution 
is cross-category.



\mypara{Structure preservation vs.~breaking}
Structure-preserving modeling and editing of man-made objects has attracted much interest in computer 
graphics~\cite{mitra2013}. This type of modeling is naturally performed only over shapes which belong to 
the same category since it is the shared structures among these shapes that are to be discovered and 
preserved. In our work however, we seek functionally plausible offsprings which may not preserve prominent 
structures of their parents. Specifically, structure breaking is possible since our functional plausibility 
score is not always positively correlated with structure preservation --- it does not account for structural 
constraints such as symmetry. As well, the composition of part groups does not need to respect symmetry, 
e.g., one armrest of a chair can form its own part group.
Our evolution is not designed to be strictly functionality-preserving either, as the hybrids can typically
break some aspects of the functionalities of their parents.

\mypara{Functionality analysis}
Recently, there have been increasing interests in studying 3D shapes from a functional 
perspective~\cite{hu2018star}. Existing works on this topic have all focused on characterizing,
comparing, or categorizing 3D objects based on their functions, which are typically inferred from
their geometry and interactions with other objects or agents~\cite{kim2014,savva2014,hu2015,hu2016,pirk2016,hu2018iconnet}.
We are not aware of previous works on functionality-aware generation or synthesis of sets of 3D objects.
Our work utilizes the \emph{category} functionality model developed by Hu et al.~\cite{hu2016}.
However, the model has to be localized to allow functionality partial matching, which is essential when defining
the plausibility of hybrid shapes which do not belong to any known functional categories.

Note that, as an application of their functionality model, Hu et al.~\cite{hu2016} were able to produce some
limited forms of model hybrids. However, the parent shapes are given and the modeling is manual, 
with functionality prediction results guiding the user in choosing where hybridization can occur. Most importantly, 
their hybridization is designed to \emph{preserve} the functionality of the parent shape.
%

\mypara{Cross-category part replacement}
%
The modeling tool developed by Zheng et al.~\cite{zheng13} does allow a specific form of 3D part sub-structures to be 
transplanted across objects categories. The key difference to our method is that their substructure replacements 
still \emph{preserve} the objects' overall structures, i.e., there is no structure breaking. Also, the detected and 
replaced substructures are rather specific: a symmetric arrangement of three parts called an \textsc{sFarr}. In contrast, 
our tool for cross-category modeling extracts functional properties of 3D objects from a more generic analysis,
and with structure breaking, the models resulting from part crossover exhibit much greater generality and 
versatility (Fig.~\ref{fig:zheng_comparison}). 

More closely related to our work is the modeling tool by Fu et al.~\cite{fu2017} which 
generates part assemblies based on \emph{affordance constraints} provided by a human pose.
Specifically, their method first identifies groups of candidate shapes which provide affordances compatible with 
the input pose, and then recombines them, possibly crossing object categories, to form a coherent 
composite shape. In contrast, our functionality analysis is entirely based on shape geometry. More importantly, 
our model evolution operates on 3D shape collections and leads to the generation of a larger volume and variety 
of functionally plausible hybrids, not just one model to fit a particular human pose, 
as demonstrated in Section~\ref{sec:results}. Finally, our modeling tool can also take on user constraints,
but in the form of functionality labels for the target shapes; such constraints are more intuitve, more natural, and 
much easier to specify from a user's perspective compared to input human poses.

%

%% file: evolution.tex

\section{Evolution Setup}
\label{sec:setup}

In this section, we describe the setup of the evolution, including the input and output of the method, and the pre-processing of the input.


\mypara{Input and output}
The input to our functionality-aware modeling tool, which serves as the initial population for the evolution, is a set of 3D shapes. Each shape is required to come with a fine-grained segmentation into meaningful parts and contact points defined for each part. The input shapes also need to be roughly aligned in a consistent manner. However, there are no strong requirements on the shape quality, e.g., shapes can be given as polygon soups, since the computations performed by our method do not require watertight meshes. 

The goal of the evolution is to stochastically create a large variety of shapes from a small set of parent shapes, according to guidance from functionality models and possibly the user. The input sets that we use in our experiments typically have less than $20$ shapes. The additional work of segmenting the input shapes is leveraged by the method for creating a large and diverse set of offspring. Iteratively, we perform crossover operations on selected pairs of shapes from the population to produce potential offspring. After several iterations, the output of the evolution is a population of shapes, where each shape combines parts coming from multiple parent shapes, possibly from distinct categories.  
Fig.~\ref{fig:overview} gives an overview of the evolution, with more details in Section~\ref{sec:evolution}.
%

\mypara{Shape representation}
The input shapes are given as triangle meshes where the triangles are grouped into parts. For each pair of adjacent parts, \emph{contact points} indicate how the parts connect to each other. As illustrated in Fig.~\ref{fig:contact}, our method considers three types of contact points, which connect two parts with four, two, or a single contact point.
These types of contact points cover the majority of part connections observed in our dataset and allow us to appropriately position parts by defining automatic connection rules. We extract the contact points in a semi-automatic manner. For each pair of parts, we detect the pair of closest points on each part and take the midpoint of these points as the contact point. Then, if needed, the user can manually specify additional contact points or adjust them manually. In our dataset, about half of the contact points were computed automatically.

Given all the contact points, a shape is abstracted as a \emph{relation graph}, where each part corresponds to a node in the graph, and two nodes are connected by an edge if the parts possess at least one contact point in common. Examples of relation graphs are shown in Fig.~\ref{fig:relation_graph}. The graphs capture the structure of the shapes and guide the placement of parts during evolution.

\input{figures/contact}

\input{figures/relation_graph}

\mypara{Functionality labels}
In our work, we employ two types of \emph{functionality labels}: (i) \emph{part labels}, which describe the functionality of a part, e.g., \emph{rolling} for a wheel or \emph{sitting} for a chair seat; (ii) \emph{category labels}, which denote the functionality of an entire category of shapes, e.g., \emph{chair} for shapes that can be used as chairs. To avoid confusion between these two types of labels, we call the former \emph{label} of a part and the latter only the \emph{category} of the shape.

Before starting the evolution, we label the parts of each shape with functionality labels. These serve as constraints for the evolutionary operations. To obtain the functionality labels for shapes of known categories, we first predict \emph{functional patches} over the initial population of shapes with the functionality models of Hu et al.~\cite{hu2016}. The prediction provides a weight field over the input point cloud, which defines the probability that each point belongs to a patch that provides a certain functionality, such as \emph{sitting} or \emph{leaning}. These patches associated with a functionality are called \emph{proto-patches}. More details about the functionality models and prediction are given in Section~\ref{sec:function}.  
%
%
%

Given the weight fields predicted for each proto-patch, we assign to a part the label of the proto-patch that has the highest sum of weights in the part. If the sum of weights is below a threshold of $0.5$ for all the functional categories (where the weights for a proto-patch sum up to $1$), then we leave the part unlabeled.
If the input shape belongs to an unknown category, the user can manually annotate its parts with custom functionality labels, e.g., \emph{rolling}. Note that some parts may remain unlabeled. We keep track of symmetries among shape parts by storing them in a list. In our work, symmetries are manually specified by the user, but any existing method for automatically detecting part symmetries can be incorporated into our pipeline~\cite{mitra12}.

Moreover, we assign each shape to a single functional category. Then, when shapes are evolved from two parents, they receive all the categories from their parents. The categories indicate which functionality models should be used for evaluating the functionality of the shapes, which we describe in more detail in Section~\ref{sec:evolution}.

%

\input{figures/part_group}

\mypara{Part groups}
Shape evolution is executed by crossover operations defined at the \emph{part group} level, where each part group consists of one or more parts from a shape in the population. We create multiple candidate part groups for each shape by enumerating possible combinations of its parts. We use a heuristic based on the part labels to form the part groups, which helps us to avoid the combinatorial explosion of enumerating all possible part groups. 

Specifically, we first group together adjacent parts with the same functionality label, which form base groups from which we build the final part groups. For each of these base groups, we create several part groups by adding different combinations of parts that are adjacent to the base group, with a breadth-first search. Note that the expanded parts need to be either all unlabeled or share a same label, which can be distinct from the initial label. A base group by itself is also considered a valid part group.
Part groups created in this manner are restricted to one or two functionality labels,
and thus form meaningful structures for exchanging and recombining the functionalities of the input shapes. We also allow symmetric parts to form base groups for further enumeration, implying that part groups can be disconnected.
Finally, individual symmetric parts can also form the basis of part groups, so that \emph{structure breaking} is allowed in the evolution. Fig.~\ref{fig:part_group} shows examples of part groups created for two different types of base groups of a shape. 

%


\section{Shape Evolution}
\label{sec:evolution}
We now describe our shape evolution framework, discussing first the evolutionary operations and then describing our method for constrained evolution. Scoring mechanisms which control the flow of the evolution will be the focus of Section~\ref{sec:function}. 

\mypara{Evolutionary operations}
A crossover is defined between two part groups $g_A$ and $g_B$, anchored on shapes $\mathcal{S}_A$ and $\mathcal{S}_B$, respectively. The crossover results in two possible offsprings. In one offspring, $g_A$ is replaced by $g_B$ on shape $\mathcal{S}_A$. In the other, $g_B$ is replaced by $g_A$ on shape $\mathcal{S}_B$. 
We also allow in some cases $g_A$ (or $g_B$) to be the null set, so that one of the offsprings would be $\mathcal{S}_B$ with $g_B$ deleted and the other is the result of inserting $g_B$ into $\mathcal{S}_A$. 
We call these two different types of crossover \emph{part group exchange} and \emph{part group insertion}.
Deformation of part groups may be necessitated after a crossover to fulfill geometric constraints. 
Note that we do not define a \emph{part group removal} operation since we
start the evolution with relatively simple shapes that we evolve into
more complex ones, so that removal of functionality is not needed.
Fig.~\ref{fig:placement1} illustrates examples of part group exchange, while Fig.~\ref{fig:insertion} illustrates insertion.

\input{figures/placement1}

\mypara{Part group exchange}
Suppose without loss of generality that we are replacing $g_B$ with $g_A$ in shape $\mathcal{S}_B$. We first perform an initial alignment of $g_A$, so that we can use spatial proximity to establish a correspondence between the contact points of $g_A$ and contact points that previously connected to parts in $g_B$.
Then, we refine the placement of $g_A$ so that corresponding contact points are brought into contact with each other, although we do not explicitly merge the triangle meshes that represent the parts.

For the initial alignment, we describe $g_A$ and $g_B$ with axis-aligned bounding boxes. We translate $g_A$ so that the center of its box aligns with the center of $g_B$. Next, we scale the longest axis of $g_A$ to align it with the corresponding axis of $g_B$, and scale the other axes proportionally to maintain the aspect ratio of $g_A$'s bounding box. 

For the refined alignment, we match each contact point in $g_A$ to the closest contact point in $\mathcal{S}_B$, after the initial alignment. If $g_A$ and $g_B$ have different numbers of contact points $n_A$ and $n_B$, respectively, we define the match only for the $n = \min(n_A, n_B)$ closest contact points. 
Then, we derive the transformation that best aligns the matching contact points. In the following operations, only points that were matched are considered. First we derive a translation that aligns the centroid of the contact points in $g_A$ to the centroid of the corresponding contact points in shape $\mathcal{S}_B$. Next, we scale $g_A$ by the average of the scalings needed to align each contact point in $g_A$ to its matching contact point in $\mathcal{S}_B$. This provides a transformation that best aligns the part groups in a least-squares sense, according to a term which is the sum of squared errors of all contact points, similarly to the optimization proposed for part placement by Kalogerakis et al.~\cite{kalogerakis2012}. Note that, since the input shapes are roughly pre-aligned, the transformation does not include a rotation. 



The two steps used in part placement are illustrated in the bottom row of Fig.~\ref{fig:placement1}. If the refined alignment is suboptimal, meaning that the distance of any of the contact points to its closest contact point in the other part group is too large, according to a threshold, then we revert back to the initial alignment. This provides a more meaningful part placement as shown in the right column of Fig.~\ref{fig:placement1}, since in this example the refined alignment fails because some of the blue contact points are too far from the yellow ones. The proximity threshold is set as $\SI{5}{\percent}$ of the shape's bounding box diagonal.

During the refined alignment, the proportions of parts can change, which could leave some parts unrecognizable. Thus, we restore the proportions of parts that possess functionality labels, if their scaling passes a threshold. Specifically, if the scaling of the $y$-axis or $z$-axis of a part's bounding box relative to the $x$-axis is greater than an empirical factor of $3$, we restore the scale of the affected axis. 


\mypara{Part group insertion}
For inserting a part group $g$ into a shape $\mathcal{S}_A$,
we look for a region in $\mathcal{S}_A$ with a similar structure to the context of $g$ in its source shape and place $g$ in this region. Specifically, we record the relative position (translation vectors) of $g$ to all of its adjacent parts in its source shape that have functionality labels, where we denote the set of labels of the adjacent parts as $\mathcal{L}_\mathrm{adjacent}$. Next, we locate a region in $\mathcal{S}_A$ which has a similar relative position to the same set of labels $\mathcal{L}_\mathrm{adjacent}$. This is formally defined as the location in $\mathcal{S}_A$ with the average of translation vectors that is the closest to the average recorded in the source shape. Only the labels in $\mathcal{L}_\mathrm{adjacent}$ that exist in shape $\mathcal{S}_A$ are considered in the average. We choose this location as the position to insert $g$.
An example of insertion is shown in Fig.~\ref{fig:insertion}.
Note that, if the location in $\mathcal{S}_A$ is already occupied by a part group $g_A$, then we perform a standard crossover to exchange $g_A$ for $g$.



%

\input{figures/insertion}

\input{algorithms/shape_evolution}

\mypara{Shape evolution}
Shape evolution is performed with crossover operations and is guided by user preferences, including preferred functionality labels and preferred offspring shapes.
Starting from an initial population $\mathcal{G}_0$, the evolution iteratively generates offspring shapes in generation $\mathcal{G}_i$ from shapes in the previous generation $\mathcal{G}_{i-1}$. All the generated shapes preserve the user-selected functionality labels specified by a set $\mathcal{L}_\mathrm{user}$. The evolution stops when a preset maximum number of iterations $i_\mathrm{max}$ is reached. 

During the generation of $\mathcal{G}_i$, the evolution considers all possible pairs of shapes, $\mathcal{S}_A$ and $\mathcal{S}_B$ ($\mathcal{S}_A \neq \mathcal{S}_B$), from $\mathcal{G}_{i-1}$ as parent shapes to generate offspring shapes. If all the user-selected functionality labels are already present in the parent shapes, we only perform crossovers that exchange part groups. If missing functionality labels are detected, we insert a part group that possesses the missing functionality into the candidate shape. Then, we apply a diversity selection to ensure that the generated shapes are geometrically \emph{diverse}. Also, we use functionality partial matching, which is discussed in Section~\ref{sec:function}, to compute the functional plausibility scores for each offspring shape, and sort the shapes in the current generation according to the descending order of their scores. Finally, after a generation $\mathcal{G}_i$ is produced, we allow the user to select preferred offspring shapes as parent shapes for the next generation. 
The flow of our shape evolution is described in Algorithm~\ref{algorithm:shape_evolution}. More details about 
the constraints used in the evolution are discussed below.


\mypara{User-in-the-loop modeling}
The user can control the evolution with two guiding mechanisms: setting constraints on the functionality part labels, and selecting preferred offsprings for further evolution.
For constrained modeling, the user selects a set of part functionalities $\mathcal{L}_\mathrm{user}$ that should appear in the offspring shapes, and the evolutionary process ensures that the shapes generated by crossovers possess all of the specified labels. In practice, the user selects the labels from a list of the labels that appear in the input population. 

%

For satisfying the user constraints, we verify if all the shapes
possess parts with the functionality labels listed in the constraints.
If a shape $\mathcal{S}_A$ does not possess all the labels, we insert the missing labels
through additional crossover operations. Specifically, for each missing functionality in $\mathcal{S}_A$, we randomly select a part group $g$ that possesses the missing functionality, from the pool of all part groups derived from the parent shapes. Next, we add the part group to $\mathcal{S}_A$ with one of two operations: (i) exchanging $g$ for a part group $g_A$ which is unlabeled or has a label that is not in the set of constraints; (ii) inserting $g$ into $\mathcal{S}_A$ with part group insertion. 
We remark that insertion operations are mainly used to add missing functionality to offsprings, as otherwise they could be applied ad infinitum in an unconstrained setting.
If the necessary part group insertions cannot be performed, then the candidate shape is not retained. 

After the set of evolved shapes had been created, it can be
further filtered with additional validity checks. In our implementation,
we include a diversity selection step.

\mypara{Diversity selection}
Given that several of the offspring shapes can be geometrically similar if they were created from similar part groups, we present to the user selected shapes that are diverse in terms of their geometry. Specifically, we compute the geometric similarity between all pairs of shapes according to the light field descriptor (LFD)~\cite{chen03}, which gives an indication of the global similarity of shapes. Next, we perform farthest point sampling according to the LFD distances, to keep only the top $\SI{50}{\percent}$ most distinct shapes. These shapes are presented to the user according to a ranking that takes into account their plausibility from a functional perspective, as described in the next section.

%% file: figures/contact.tex
\begin{figure}[t]
    \centering
    \includegraphics[width=\linewidth]{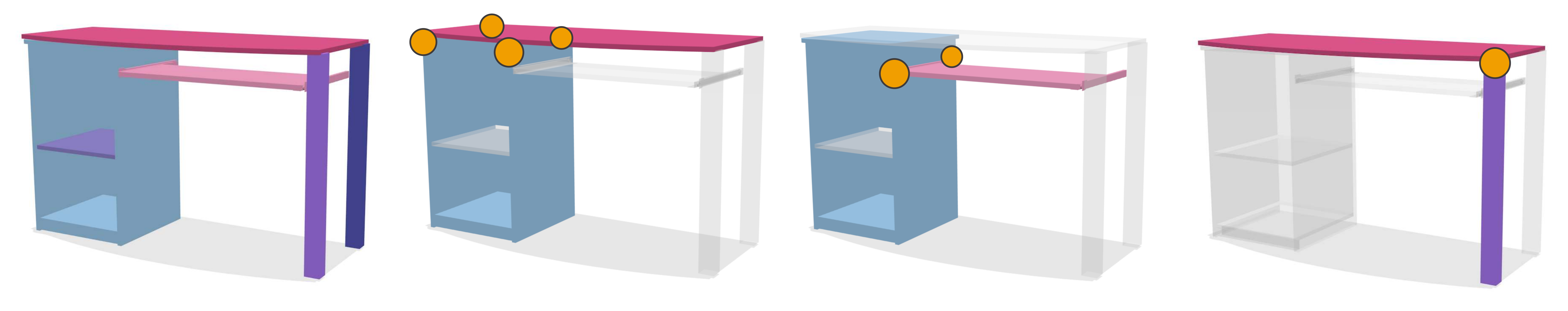}
    \caption{Examples of part pairs with different numbers of contact points: four, two, and one contact point.}
    \label{fig:contact}
\end{figure}

%% file: figures/relation_graph.tex
\begin{figure}
    \centering
    \includegraphics[width=0.90\linewidth,trim={85 85 85 95},clip]{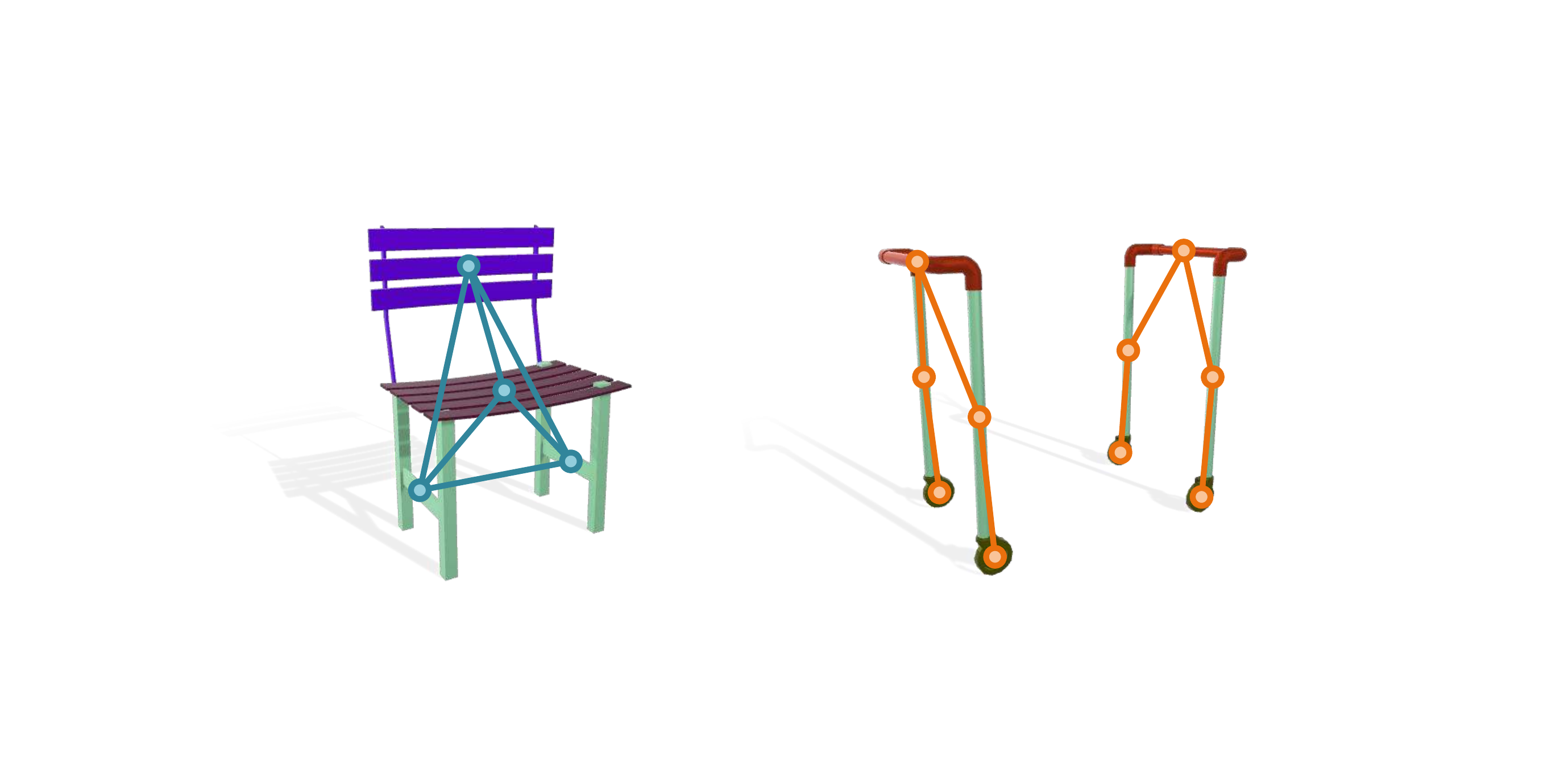}
    \caption{Examples of relation graphs for two shapes: a connected graph (left) and a disconnected graph (right).}
    \label{fig:relation_graph}
\end{figure}

%% file: figures/part_group.tex
\begin{figure}[t]
    \centering
    \includegraphics[width=\linewidth]{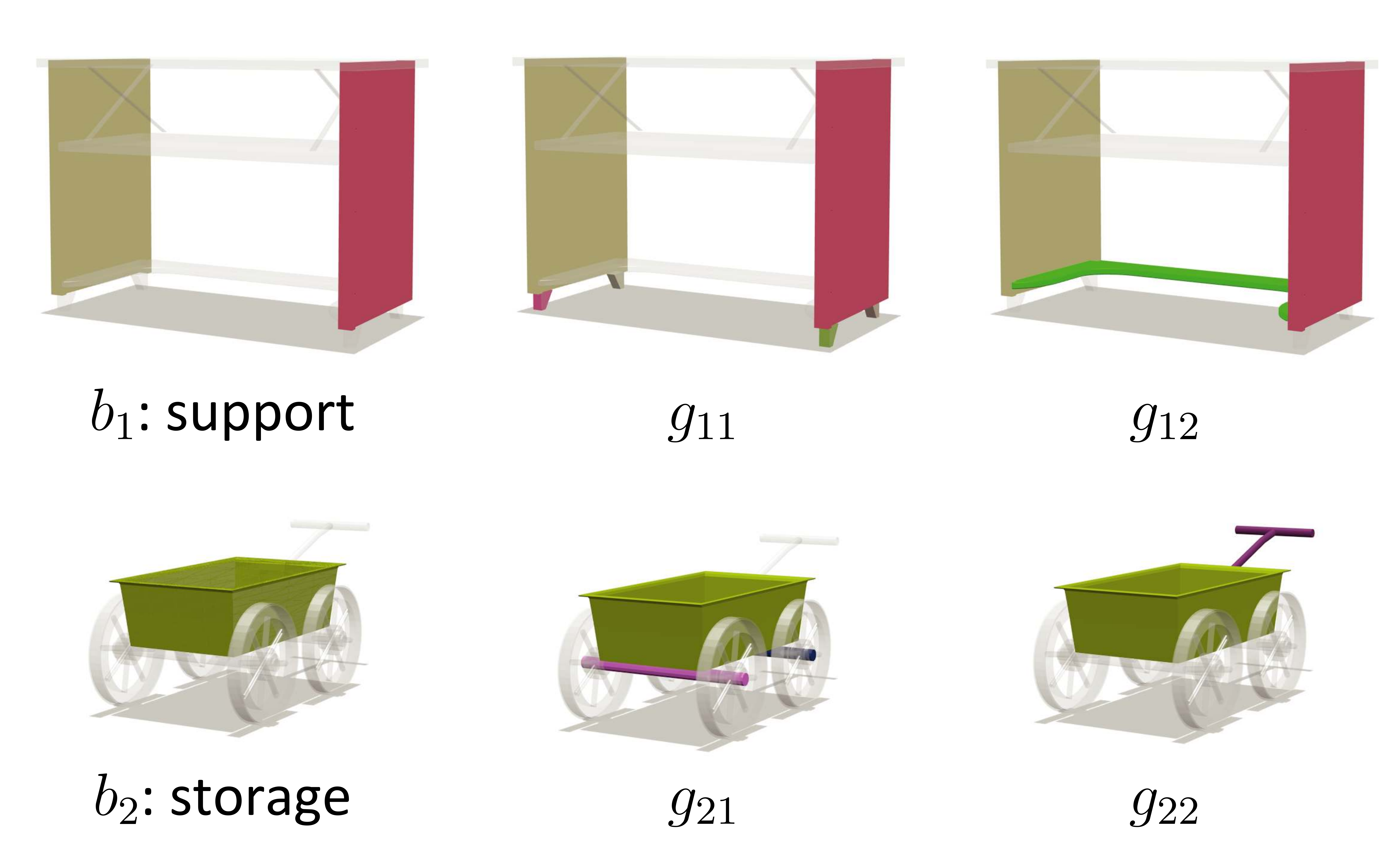}
    \caption{Examples of part groups derived from base groups. Given base group $b_1$, with functionality label \emph{support}, we derive part groups $g_{11}$ and $g_{12}$ by adding the colored parts to $b_1$, where $b_1$ is formed by symmetric and disconnected parts. Similarly, we derive $g_{21}$ and $g_{22}$ from $b_2$.}
    \label{fig:part_group}
\end{figure}

%% file: figures/placement1.tex
\begin{figure}
    \centering
    \includegraphics[width=\linewidth]{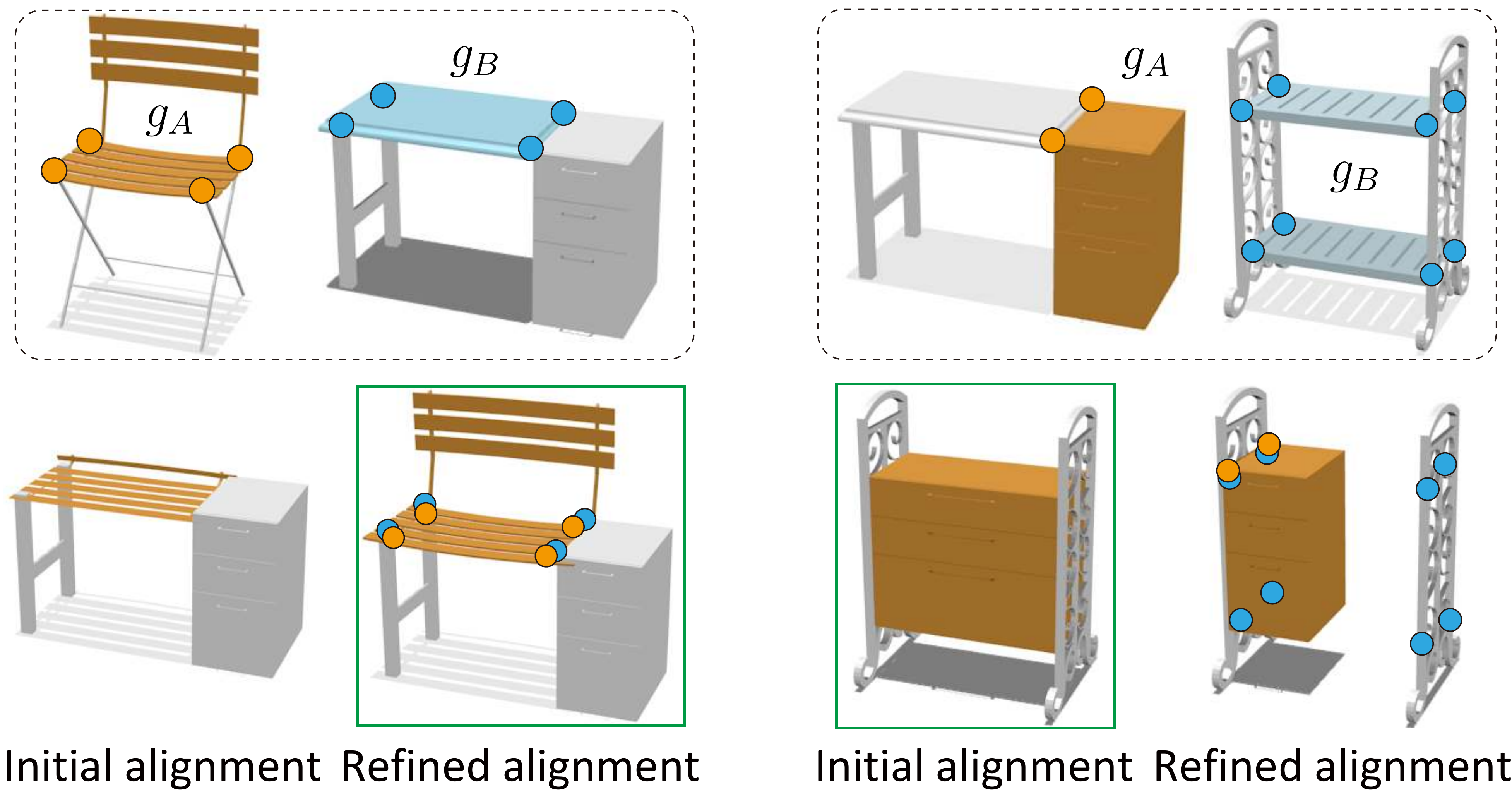}
    \caption{Two examples of part group exchange. After a crossover that replaces the part group $g_B$ by $g_A$, we perform an initial placement of $g_A$ based on bounding box alignment. This is followed by a refinement deformation that aligns contact points (blue and orange circles).
    The refinement may be reverted back if some of the contact points are not brought into proximity (right column example).}
    \label{fig:placement1}
\end{figure}

%% file: figures/insertion.tex
\begin{figure}
    \centering
    \includegraphics[width=\linewidth]{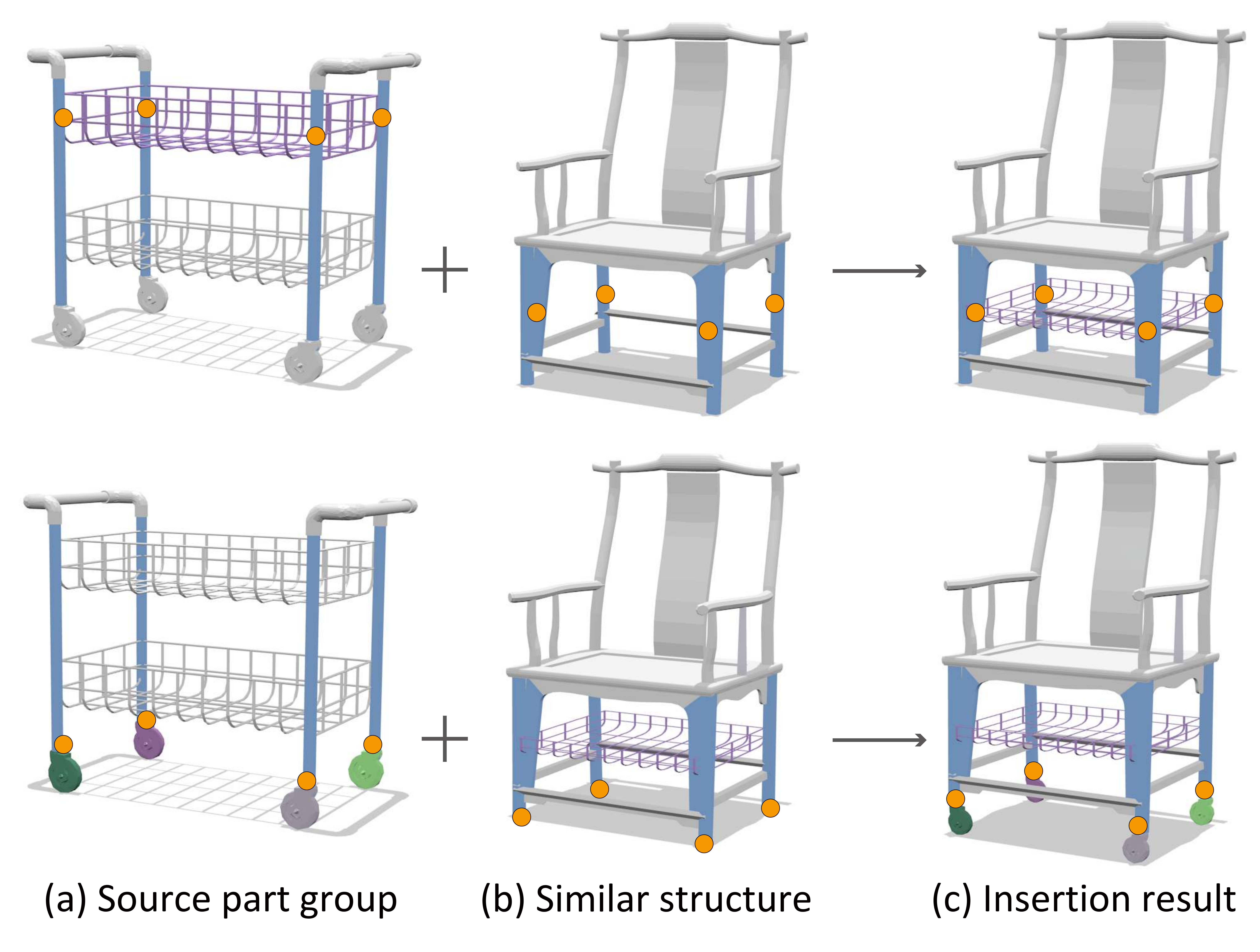}
    \caption{Two examples of part group insertion. Given a part group in (a), we find a region in the target shape in (b) with a structure similar to the context of the part group in the source shape in (a). For example, the context consists of four support structures (the legs) adjacent to the basket on the top row, or one support structure (one leg) adjacent to each wheel on the bottom row. We insert the part group in this region as shown in (c).}
    \label{fig:insertion}
\end{figure}

%% file: algorithms/shape_evolution.tex
\begin{algorithm}[t]
    \caption{Shape evolution}\label{algorithm:shape_evolution}
    \DontPrintSemicolon
    \KwIn{$\mathcal{G}_0, \mathcal{L}_\mathrm{user}, i_\mathrm{max}$ (refer to the text for symbol notation)}
    \KwOut{$\mathcal{G}_\mathrm{evolved}$}
    \SetKwProg{Fn}{function}{}{end}\SetKwFunction{FShapeEvolution}{ShapeEvolution}
    \Fn{\FShapeEvolution{\textup{\texttt{$\mathcal{G}_0$,$\mathcal{L}_\mathrm{user}$,$i_\mathrm{max}$}}}}{
        $\mathcal{G}_\mathrm{evolved} \gets \emptyset$\;
        $i \gets 1$\;
        \While{$i \leq i_\mathrm{max}$}{
            $\mathcal{G}_i \gets \emptyset$\;
            \ForEach{$\mathcal{S}_A \in \mathcal{G}_{i-1}$}{
                $\mathcal{L}_A \gets$ functionality labels in $\mathcal{S}_A$\;
                $\mathcal{L}_\mathrm{missing} \gets \mathcal{L}_\mathrm{user} \setminus \mathcal{L}_A$\;
                \ForEach{$\mathcal{S}_B \in \mathcal{G}_{i-1}$}{
                    \If{$\mathcal{S}_A \neq \mathcal{S}_B$}{
                        \If{$\mathcal{L}_\mathrm{missing} = \emptyset$}{
                            \ForEach{$g_A \in$ part groups of $\mathcal{S}_A$}{
                                \ForEach{$g_B \in$ part groups of $\mathcal{S}_B$}{
                                    \If{$g_A$ is unlabeled or has no functionality label in $\mathcal{L}_\mathrm{user}$}{
                                        $\mathcal{S}_\mathrm{offspring} \gets$ exchange $g_A$ on $\mathcal{S}_A$ for $g_B$\;
                                        $\mathcal{G}_i \gets \mathcal{G}_i \cup \{\mathcal{S}_\mathrm{offspring}\}$\;
                                    }
                                }
                            }
                        }
                        \If{$\mathcal{L}_\mathrm{missing} \neq \emptyset$}{
                            \ForEach{$g \in$ part groups of $\mathcal{S}_B$}{
                                \If{$g$ has a functionality label in $\mathcal{L}_\mathrm{missing}$}{
                                    $\mathcal{S}_\mathrm{offspring} \gets$ insert $g$ into $\mathcal{S}_A$\;
                                    $\mathcal{G}_i \gets \mathcal{G}_i \cup \{\mathcal{S}_\mathrm{offspring}\}$\;
                                }
                            }
                        }
                    }
                }
            }
            $\mathcal{G}_i \gets \texttt{DiversitySelection($\mathcal{G}_i$)}$\;
            $\texttt{scores} \gets \emptyset$\;
            \ForEach{$\mathcal{S} \in \mathcal{G}_i$}{
                $s \gets \texttt{FunctionalityPartialMatching($\mathcal{S}$)}$\;
                $\texttt{scores} \gets \texttt{scores} \cup \{s\}$\;
            }
            $\mathcal{G}_i \gets \texttt{Sort($\mathcal{G}_i$,scores)}$\;
            $\mathcal{G}_i \gets \texttt{UserSelection($\mathcal{G}_i$)}$\;
            $\mathcal{G}_\mathrm{evolved} \gets \mathcal{G}_\mathrm{evolved} \cup \mathcal{G}_i$\;
            $i \gets i + 1$\;
        }
        \Return{$\mathcal{G}_\mathrm{evolved}$}\;
    }
\end{algorithm}

%% file: function.tex
\section{Functionality Analysis and Scoring}
\label{sec:function}

In this section, we explain how we evaluate the functional plausibility of the evolved shapes.


\mypara{Category functionality model}
%
%
We base our functionality analysis on the model of Hu et al.~\cite{hu2016}. We provide a short description of this model here for completeness, but refer the reader to the original paper for more details~\cite{hu2016}. Each model captures the functionality of an entire category of shapes with a set of \emph{proto-patches} and their geometric descriptors. A \emph{proto-patch} is a generalization of patches that support a specific type of interaction, e.g., handles for grasping. The model stores the proto-patches for all the interactions that are needed to enable the functionality of a category, e.g., proto-patches of handles, baskets, and wheels of supermarket carts. Each shape is described as a set of points, and each proto-patch is represented as a weight field defined over the points. The weight of a point represents the likelihood that the point belongs to the specific proto-patch. 

Besides the proto-patches, the complete model stores the typical values of geometric descriptors of points that appear on the proto-patches, and also binary descriptors defined between pairs of points belonging to two different proto-patches.
The proto-patches in the functionality models allow us to \emph{localize} the functionality of the shapes. We use this localization to infer the functionality labels of parts, as described in Section~\ref{sec:setup}.

The models are learned from a set of training shapes given in contexts that describe the functionality of the shapes~\cite{hu2016}. Given an unknown shape, we can use the model of a category to predict proto-patches on the shape, and then compare the descriptor values of the proto-patches to those stored in the model, to determine how well the shape satisfies the functionality of the category (its \emph{functionality score} for the category). Note that we use functionality models for \emph{categories} of shapes since these can be learned from existing data~\cite{hu2016}, while generic models of functionality typically depend on ad hoc geometric rules that do not generalize.
%

\mypara{Functionality partial matching}
%
During the shape evolution, we can use the category models to compute the functionality scores of the offsprings, to obtain an indication of their functional plausibility. However, given that the parent shapes are of different categories, we may generate cross-category offsprings that do not fit well a single category. Moreover, one of the goals of the evolution is to generate shapes with surprising functionality, possibly mixing the functionality of different categories. Thus, we need to take multiple categories into account when evaluating an offspring.

We stipulate that a cross-category shape is functionally plausible as long as it partially supports the functionality of multiple categories. Thus, we arrive at the notion of \emph{functionality partial matching}, where we derive the functional score of a shape by aggregating the scores of partial matchings of parts of the shape to different category functionality models. Computing such partial matching requires addressing three challenges. First, the functionality scores of partial shapes derived from the input shape need to be computed efficiently for a given category model. Second, the functionality scores of different categories need to be normalized, so that their scores can be mutually compared. Third, the scores of multiple categories need to be aggregated in a meaningful manner to provide the overall partial matching score. We discuss how we address these three challenges as follows.


\mypara{Functionality score of partial shapes}
Given a category model $\mathcal{M}$ and a shape $\mathcal{S}$ to be evaluated, we enumerate subsets of shape parts, including subsets of size $1$, $2$, and so on. Next, we consider each subset as an individual shape and evaluate its functionality according to $\mathcal{M}$. Given that we are interested in the partial shape that best supports the functionality of the category, we search for the subset of parts with highest functionality score. To avoid enumerating and evaluating all possible subsets of parts, we perform the enumeration more efficiently with a beam search approach described as follows. 


\input{figures/beam_search}

Since we experimentally observed that the full objects have the highest functionality
scores in a large number of instances, we perform a \emph{reverse beam
search}, where we start with the entire shape and evaluate subsets
obtained by
removing parts of the shape one at a time. In the search, we keep track of the $w$ subsets of parts with the top functionality scores computed up to the current iteration, and continue expanding recursively only these subsets, where $w$ is the beam width. We use $w = 2$ in our experiments, but always expand all of the nodes in the first level of the search. To expand a node in the search, we generate each subset obtained by removing one part from the node's set of parts. The search stops when the $w$ nodes with higher scores were expanded, or when no further parts can be removed from any of the $w$ nodes.  Fig.~\ref{fig:beam_search} illustrates the beam search.


\input{figures/func_space}

In addition to having a high score, a partial shape created during the search should satisfy three constraints: 

(i) All parts of the partial shape should be connected. Since we use a relation graph to represent each shape, the part-wise connectivity check can be reduced to checking the graph connectivity. For example, in Fig.~\ref{fig:relation_graph}, the relation graph on the left represents a shape with all parts connected, while the one on the right represents a disconnected shape. 

(ii) The partial shape should be physically stable. Similar to the idea of static stability of shapes introduced by Fu et al.~\cite{fu2008upright}, we verify whether the center of mass of the partial shape falls within the polygon formed by all the ground-touching points of the shape. In more detail, we first calculate the center of mass of the shape by averaging the center points of the bounding box diagonals of all parts of the shape. We also compute the convex polygon formed by all \emph{ground-touching points} of the shape. To decide whether points are touching the ground, we find the minimum $z$-coordinate $z_\mathrm{min}$ among all points in the shape and mark all points that have $z$-coordinate smaller than $z_\mathrm{min} + \SI{1}{\percent}$ of the bounding box diagonal as ground-touching points. Then, the shape is considered as physically stable if the projection of the center of mass on the ground is inside the convex polygon. During the search, partial shapes are allowed to be unstable, since this may only be temporary while the final nodes of the search will be stable. Thus, we 
keep the top stable shapes in addition to the top $w$ partial shapes, in case none of the top scoring shapes is stable.

(iii) The partial shape should have an adequate functional space to support 
a function, e.g., a human may not sit on a seat if the seat is obstructed by other parts. We use the functional space extracted by the functionality model of Hu et al.~\cite{hu2016} to verify that the space needed to perform the function associated to the label is free. A few examples are shown in Fig.~\ref{fig:func_space}.



\input{figures/score}

\mypara{Functionality score normalization}
The functionality scores computed with the model of Hu et al.~\cite{hu2016} are based on the similarity between the input shape being evaluated and training shapes in the category of the model. Since the functionality models are trained separately per category, scores for different categories are not directly comparable. 
To address this limitation and obtain scores that are comparable across categories, we normalize the scores for each category according to the training data.

Given a category model, we compute the scores for all the shapes in the training data. Next, we compute the cumulative distribution of the scores of shapes inside the category, and also of the scores of shapes outside the category, obtaining cumulative distributions $\mathcal{D}_1$ and $\mathcal{D}_2$, respectively. Given an unknown query shape, we compute its score with the category model, and compute its normalized score as $w_1 \times p_1+w_2 \times p_2$, where $p_1$ is the probability of the shape's score according to $\mathcal{D}_1$, and $p_2$ according to $\mathcal{D}_2$. The probability $p_1$ is an approximation of the percentage of shapes inside the category that have scores lower than the query, and represents the probability of the query being inside the category. Similarly, $p_2$ represents the probability of the shape not being outside the category. 
Thus, the query is more likely to belong to the category when both $p_1$ and $p_2$ are large, which can be captured by a weighted sum of these two probabilities. The weights $w_1$ and $w_2$ are defined as the percentages of shapes inside and outside the category over all the shapes in the training data. They indicate the reliability of the two distributions, respectively.
Fig.~\ref{fig:score} shows one example to illustrate how the scores are normalized and more meaningful than the original scores.

\mypara{Functional plausibility score}
Given the best functionality scores of a shape computed for each of the relevant category models through partial matching and normalized as described above, we integrate the scores into a single number that indicates the shape's \emph{cross-category functional plausibility} by simply taking the maximum of the scores. Although different manners of integrating the scores are possible, the maximum score indicates the best partial functionality that the shape possesses, and thus sets a lower bound for the functionality of the shape. An alternative approach such as the sum of scores would be biased by the categories that possess low scores because the shape does not support their functionality. However, such low-score categories do not necessarily indicate that the shape is not functional. One limitation of the maximum is that it does not indicate whether a shape has multiple functionalities. To take into account such multi-functional shapes, the score described next can be used as an alternative to the plausibility score. 
%


\mypara{Multi-functionality score}
The goal of the multi-functionality score is to detect shapes that combine multiple, known functionalities in unusual manners. This score can help in measuring the ``surprise factor'' or unpredictability of an offspring.
We define the score as the number of different functionality categories that are partially supported by the shape. Specifically,
we count the number of functionality categories for which the shape has a high score, determined by whether the maximum of the functionality partial matching probability for the category is above a threshold $\theta$. We use $\theta = 0.9$ in all of our experiments. In Section~\ref{sec:results}, we show how the multi-functionality score can be used in place of the plausibility score to rank shapes, when the goal is to select hybrid shapes that support multiple functionalities.

\mypara{Trade-off between user and machine time}
The two scores described above provide a good indication of the functionalities that a shape can support. In general, the evolved shapes with high scores are plausible and thus the users can focus on selecting shapes based on their design interests. On the other hand, the computation of the scores via the functionality partial matching relies on an expensive search procedure, implying that the time to evolve a single generation can be considerably long. Thus, we also provide the user with an option to run a simplified partial matching that helps to balance between user and machine time. For each evolved shape, we compute its functionality scores based only on three sets of parts: the entire shape, and each part group coming from its two parents. This simplification speeds up the computation time considerably, at the expense that the user may have to inspect a larger number of shapes, since some plausible shapes may not be ranked correctly.

%% file: figures/beam_search.tex
\begin{figure}[t]
    \centering
    \includegraphics[width=\linewidth]{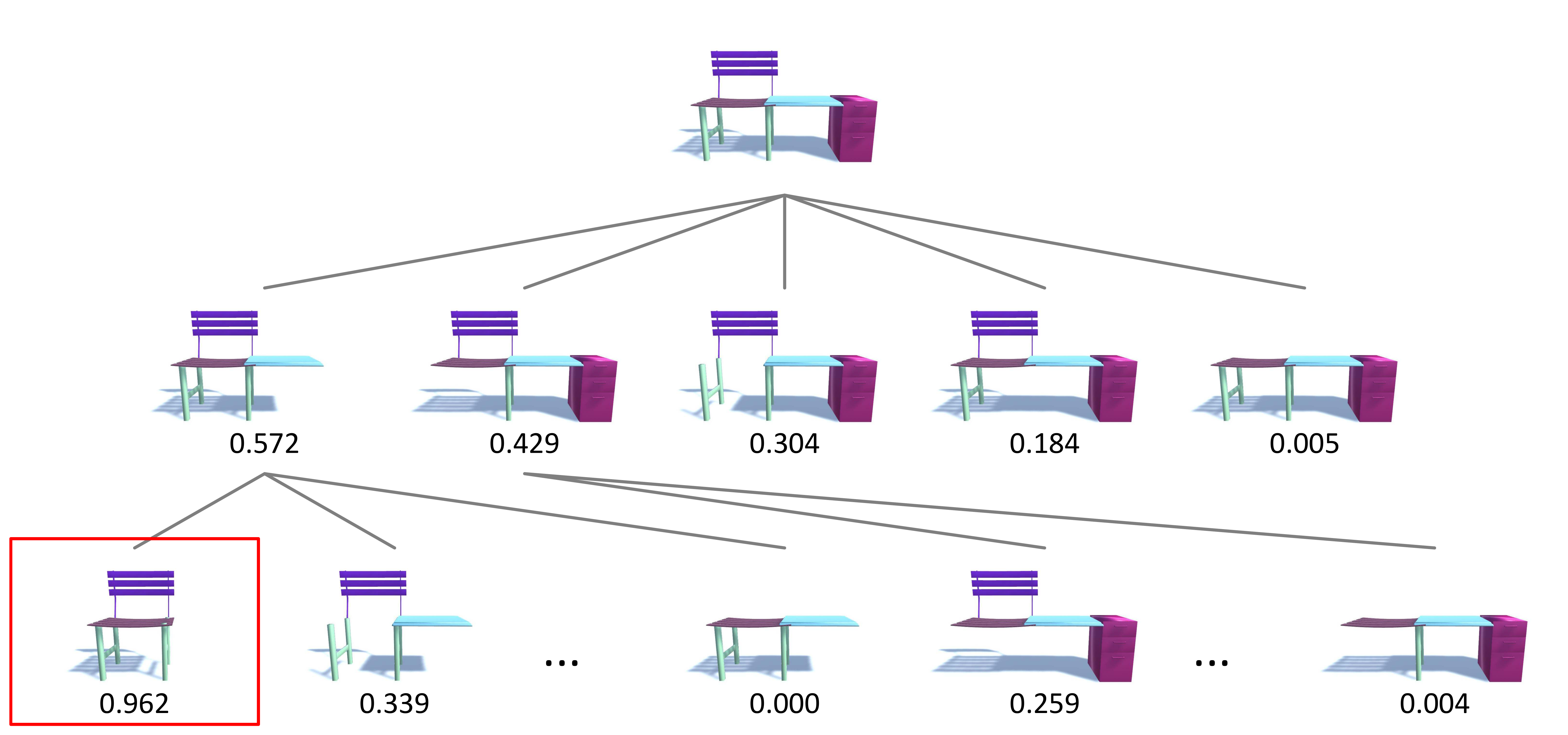}
    \caption{Beam search for functionality partial matching, where we search for the subset of parts that provides the highest score for a functional category (\emph{chair} in this example). We start with the full shape and enumerate its subsets by removing parts one at a time. We only expand a node if it is promising, i.e., it has one of the top two functionality scores, as the beam width is $2$. The node from level~$2$ highlighted in red is returned as the best partial match, since its children (omitted from the figure) have lower scores. The full search tree is shown in the supplementary material.}
    \label{fig:beam_search}
\end{figure}

%% file: figures/func_space.tex
\begin{figure}[t]
    \centering
    \includegraphics[width=0.9\linewidth]{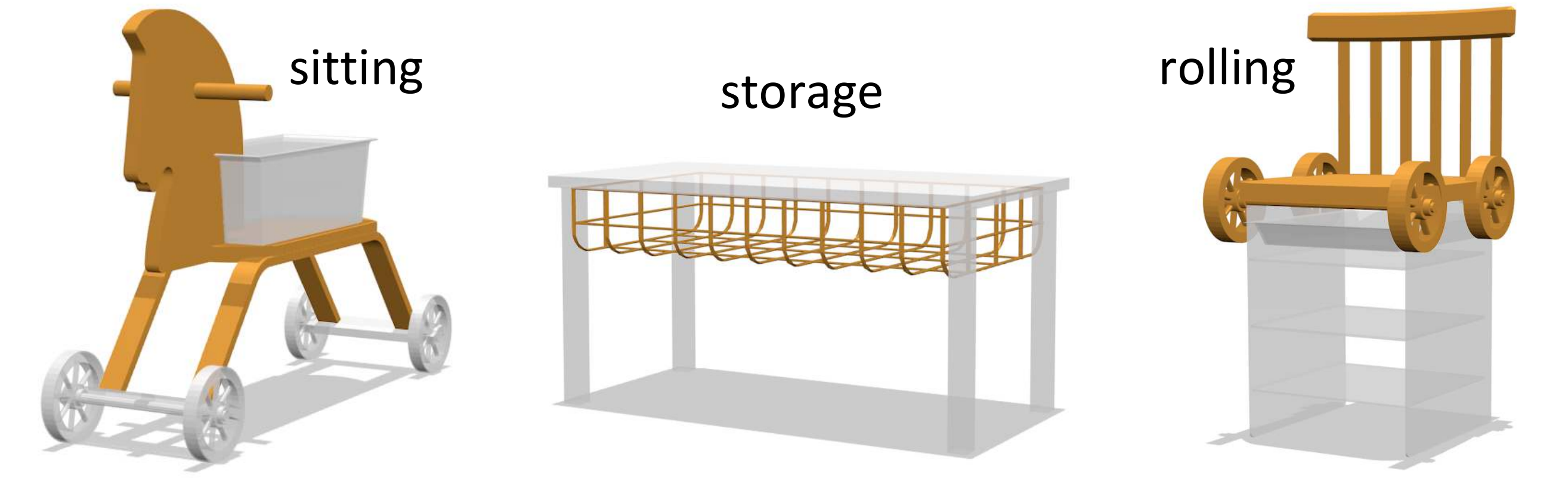}
    \caption{Subsets of parts with high partial matching score (shown in orange) which are functionally implausible due to their surrounding space being obstructed by other parts. 
    }
    \label{fig:func_space}
\end{figure}

%% file: figures/score.tex
\begin{figure}[t]
    \centering
    \includegraphics[width=\linewidth]{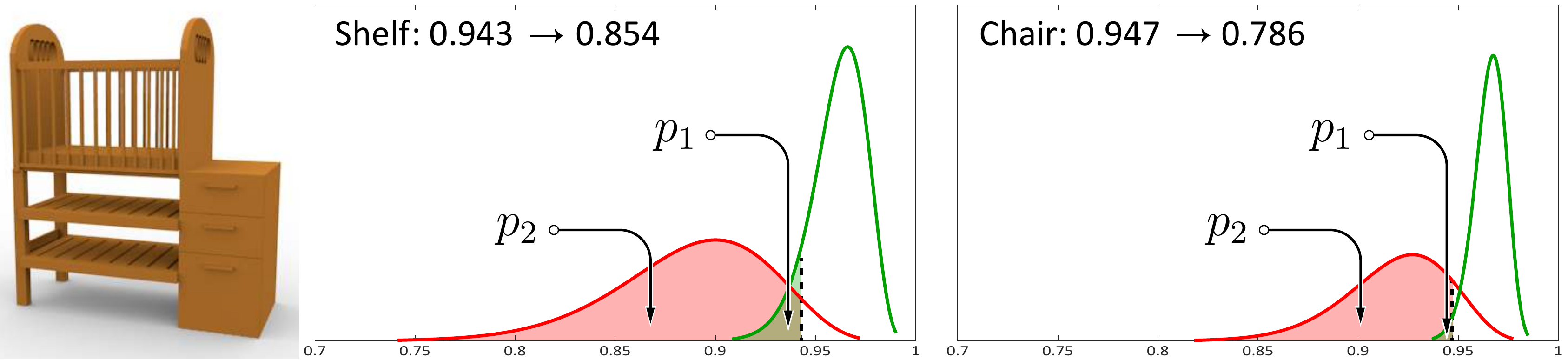}
    \caption{Normalization of functionality scores for the shape on the left. The two numbers on the top of each graph are the scores of the shape for the \emph{shelf} and \emph{chair} categories before and after normalization. The score distributions for training shapes inside/outside the corresponding category are drawn in green and red, respectively. The probabilities $p_1$ and $p_2$ represent the percentages of shapes having lower scores in the corresponding distributions. We see that, even though the original scores for the two categories are quite close, the normalized score for the \emph{chair} category becomes much smaller than that of the \emph{shelf} category, since shapes inside the \emph{chair} category have relatively high scores.
    }
    \label{fig:score}
\end{figure}

%% file: results.tex
\section{Results, Evaluation, and Applications}
\label{sec:results}

In this section, we show results of our functionality-aware model evolution under different modeling scenarios and also 
evaluate various aspects of our modeling framework. 
We also 
provide comparisons to
the modeling tools by Zheng et al.~\cite{zheng13} and Fu et al.~\cite{fu2017}. Finally, we show an application
of our modeling results to data augmentation for shape segmentation.


\mypara{Dataset and functionality models}
Our experiments have been conducted mainly on 3D objects belonging to $15$ functional categories including chairs, 
handcarts, baskets, etc. However, as we shall show, the initial population of our model evolution is not constrained 
to only include objects with known functional categories.
The $15$ functional categories have been adopted from the work of Hu et al.~\cite{hu2016}, which developed 
category functionality models for these classes. In our work, we leverage these functionality models with our 
functionality partial matching so that they can be applied to cross-category models. Note that a few categories 
such as hook, hanger and backpack do not have many parts that can be exchanged. Thus, they do not appear in 
our results, as they provide only a small number of part groups.

\input{figures/unconstrained}

\input{figures/constrained}

\input{figures/multi-func}

\input{figures/large_set}

\mypara{Parameters and statistics}
All our experiments were conducted with the \emph{same} parameter settings, as described in preceding sections. As an example of the execution time, pre-processing of an initial population of $4$ shapes takes on average $\SI{15}{\minute}$, including functional proto-patch mapping and part group creation. Then, evolving one generation with random crossovers from this population of $4$ shapes with a pool of $17$ part groups (including null part groups) and $8$ functionality labels generates $31$ shapes. On the other hand, with user constraints (preserving two functionality labels), only $6$ offspring shapes are produced. For the execution time, if a simplified partial matching is adopted, the time needed to evolve one generation is about $\SI{6}{\minute}$ on average, where evolutionary operations take about $\SI{1}{\minute}$ ($\SI{17}{\percent}$ of the computation time for a generation) and functionality partial matching takes $\SI{5}{\minute}$ (the remaining $\SI{83}{\percent}$ of the computation). The time needed to compute the functional plausibility score of a single offspring is $\SI{50}{\second}$ on average with the simplified partial matching. If a reverse beam search is employed, the time for evolving one generation from the same population is about $\SI{55}{\minute}$, where evolutionary operations take about $\SI{1}{\minute}$ ($\SI{2}{\percent}$ of the computation time) and functionality partial matching takes $\SI{54}{\minute}$ (the remaining $\SI{98}{\percent}$ of the computation). The time needed to compute the functional plausibility score of a single offspring with the reverse beam search is on average $\SI{9}{\minute}$.

The largest population we evolved includes $17$ shapes, taking on average $\SI{2.2}{\hour}$ to evolve one generation (with user constraints and the simplified partial matching).
All timings were measured on a workstation with an Intel Core i7-6700 $\SI{3.40}{\GHz}$ quad-core processor. Our implementation\footnote{Our implementation is available at the following URL: \url{https://github.com/IsaacGuan/FAME}.} is based on C\# and MATLAB and is largely unoptimized; we believe that it can be sped up significantly with refactoring.

\subsection{Evolutionary Modeling Results}

In this section, we present a variety of shape evolution results obtained with our method. If not indicated otherwise, for each figure, we show selected shapes from the evolved populations, mimicking the process where users would handpick  interesting offspring from a large and diverse set of shapes. In Fig.~\ref{fig:large_set} and Fig.~\ref{fig:rank}, we evaluate specifically the rankings produced by our method.

\mypara{Unconstrained evolution}
Fig.~\ref{fig:unconstrained} shows evolution results corresponding to an open-ended exploration, 
where the user only selected an initial population encompassing multiple functional categories, without adding
additional constraints.
We can observe that many of the evolved shapes are truly cross-category, combining the functionality of 
two or more parents, such as the movable shelves and table-shelves in the first column, or the sittable shelves 
in the second column. As the evolution advances, the complexity of the offsprings also increases, where we see a 
variety of multi-functional objects. In the second column, we see in the initial population, an object, 
the horse-shaped toy, that is not part of any of the pre-defined functional categories. By manually labeling the 
toy's parts with appropriate functionality labels such as \emph{rocking} and \emph{sitting}, our evolution can combine the 
object with other shapes to create offsprings that preserve these functionalities. Finally, we note that the functionality 
of the parent shapes is preserved in the offsprings, without obstructions to the functional spaces.

\mypara{Constrained modeling}
Evolution guided by user constraints is the modeling scenario where our tool is most effective. The user can start 
with a design goal to ``generate shapes that enable storage and transportation'', and select 
appropriate functionality labels as constraints to ensure that these functionalities appear in the offsprings. During
evolution, our tool provides concrete examples of shapes satisfying the constraints. The user can examine 
the generated shapes to select and further evolve the preferred ones.

Fig.~\ref{fig:constrained} shows a first set of results of constrained evolution, where we show three generations that were evolved from one input population. We observe how the shapes of a generation possess all of the functionalities specified as constraints, while at the same time they provide different variations on how these functionalities can be enabled, e.g., several objects that combine placement of objects (like shelves) with grasping of handles for transportation in G1 and G2, and a variety of storage furniture like baskets and shelves that can be rolled in G3, where several of the evolved objects also inherit the grasping or placement aspects from parents in G2.
%


\mypara{Structure breaking}
The ability to break structures, such as symmetries, during part composition is a unique feature of our cross-category modeling tool, which allows the evolution to introduce variations in the structure of the generated shapes. Both results in Fig.~\ref{fig:teaser} exhibit structure breaking. 
The percentage of offspring exhibiting symmetry breaking can vary considerably, e.g., $\SI{45}{\percent}$ for the set shown on the left of Fig.~\ref{fig:unconstrained}, and $\SI{6.4}{\percent}$ for the set on the right.

\subsection{Evaluation and Comparisons}

\mypara{Scalability}
We evaluate the scalability of our evolution in two aspects: the number of shapes and number of constraints involved. In Fig.~\ref{fig:multi-func}, we see that the evolution produces plausible results when requesting a small number of functionalities such as $2$ or $3$, but also a larger number such as $5$. The generated shapes are true hybrids that can serve multiple functionalities, such as the chairs in Fig.~\ref{fig:multi-func}(c) that, besides enabling sitting and leaning, also enable the placement of objects, and rolling and grasping for transportation.

Fig.~\ref{fig:large_set} presents three sets obtained by evolving the same initial population of $17$ shapes by choosing different sets of functionality constraints. The input population is larger than the input sets used in the previous experiments to assess scalability. We observe how the evolution generates large sets with hundreds of novel shapes which are all constrained by the user guidance.


\mypara{Ranking scores}
When a large number of offsprings are generated, a ranking measure is especially important so that the most plausible objects are first presented to the user. In Fig.~\ref{fig:rank}, we compare the functional plausibility and multi-functionality measures as choices for ranking the shapes from Fig.~\ref{fig:large_set}. We observe how objects with high values for the two measures appear more functional than objects with low scores. In addition, objects with high multi-functionality scores tend to combine several functionalities. Thus, these measures enable the user to save time by only inspecting the most promising prototypes.
We further evaluate the ranking measures with two user studies that we
present in the supplementary material, where we show that our
functionality scores are able to capture the multiple functionalities of
a shape as perceived by humans.


\input{figures/rank}

\input{figures/zheng_comparison}

\mypara{Comparison to Zheng et al.~\cite{zheng13}}
In Fig.~\ref{fig:zheng_comparison}, we compare our method to that of Zheng et al.~\cite{zheng13} by applying our model evolution to one of their input sets, with the same shape segmentations. Recall that their method is designed to preserve a specific type of three-part symmetric support substructure (\textsc{sFarr}) in the input and limited to six combination rules, while our method is more generic and even allows symmetry breaking. We observe that our generic hybridization/crossover approach, i.e., without explicitly modeling or enforcing any specific symmetries or support structures, can also generate the types of shapes that their method can. At the same time,
our modeling tool is able to achieve more general part recombinations and produce shapes exhibiting larger functional variations, e.g., the combination of sitting and storage that appears 
in the couch hybridized with a bench and shelf, where two different parts are attached to the supporting structure. Moreover, we remark that their method is able to generate interesting variations for this input set where all shapes have symmetric support structures, but their method is not applicable to more general shapes like the asymmetric desks that we evolved in Fig.~\ref{fig:teaser} or Fig.~\ref{fig:large_set}.

\input{figures/fu_comparison}

\mypara{Comparison to Fu et al.~\cite{fu2017}}
In Fig.~\ref{fig:compare_fu}, we compare our method to that of Fu et al.~\cite{fu2017}, which can synthesize a cross-category, functional hybrid to fulfill the affordance constraint defined by an input human pose. The comparison results demonstrate that our evolutionary modeling tool is able to produce similar hybrids, without specifying a human pose or explicit affordance constraints, as well as other hybrids which are also functionally plausible. On the other hand, an input human pose does narrow down the search for potential parent shapes and part placements; the resulting synthesized shape would more closely serve a specific target functionality.

\input{figures/data_aug}

\input{figures/data_aug_visual}

\subsection{Application to Data Augmentation}

We have shown that our functionality-aware model evolution is able to generate a large and diverse population of functionally plausible offsprings. But, from the perspective of modeling and design exploration, we should expect the functionality requirement placed on the final outcomes to be quite stringent. Hence, in an evolved population, there are usually only few shapes that are fully functional or can inspire new designs. On the other hand, data augmentation, which is aimed at boosting the performance of learning-based shape analysis schemes, is a venue where we can utilize most, if not all, of the evolved shapes from our modeling tool. We expect these shapes to serve as useful training data since they are both ``fit'', i.e., plausible, to resemble, at least in part, potential test shapes, and diverse, to provide a better coverage of the distribution of test shapes.


To demonstrate the potential of our method for data augmentation, we use the generated offspring to augment training data for one key application: shape segmentation of partial shapes. To create a set of labeled shape segmentations, we assign labels to the already segmented parent shapes, which usually constitute a small set that can be manually labeled by a human in a short time. Then, we keep the labels assigned to each part as we evolve the parents into a large set of segmented and labeled shapes. We use this data to augment the training data for shape segmentation. 

Specifically, we evolve two input populations. The first population is composed of $7$ chairs and $9$ shapes from other classes that add diversity to the set, resulting in an evolved population of $138$ hybrid chairs, since we constrain the functionality of chairs to be present in the offspring. The second population is composed of $10$ tables and $5$ shapes from other classes, resulting in $100$ hybrid tables.
For the experiments, we use PointNet~\cite{qi17} for segmenting point clouds, where we uniformly sample our shapes with $2048$ points.
To show how our data can aid in learning to segment diverse shapes, especially when partial matching is needed, we create a test set of partial shapes, where parts are randomly removed from ShapeNet~\cite{chang2015_shapeNet} shapes. We augment the ShapeNet training set used by PointNet with our training shapes composed of hybrid shapes, and evaluate the predicted segmentations on the test set of partial shapes.

Fig.~\ref{fig:data_aug} shows the results of data augmentation with increasing numbers of shapes from our training set, where we evaluate the segmentations by measuring the label accuracy. We observe how each additional batch of shapes helps to improve the accuracy on the chair set, with a gain of around $\SI{8}{\percent}$. The accuracy is also slightly improved for tables, which typically have a simpler structure. Moreover, the data points with $x = 0$ represent the accuracy obtained when only the ShapeNet training set is used for learning. We notice how the accuracy is much lower than when our more diverse shapes are used for training, demonstrating that the diverse, plausible shapes do provide additional information for the learning of the deep network. 

We confirm this reasoning by visually inspecting the results, such as the set of examples shown in Fig.~\ref{fig:data_aug_visual}. Note how the segmentations of these partial shapes have considerable errors when only the ShapeNet set is used for training. On the other hand, we obtain a much refined segmentation, closer to the ground truth, when our evolved shapes are used for data augmentation. For example, the chair-stool combo in the second row is missing the side handles, but the ShapeNet data alone still leads to a segmentation with this label. In the third row, the leg label appears below the seat, while with our data a better prediction of chair seats and back rests is obtained.

%% file: figures/unconstrained.tex
\begin{figure*}[t]
    \centering
    \includegraphics[width=0.95\linewidth]{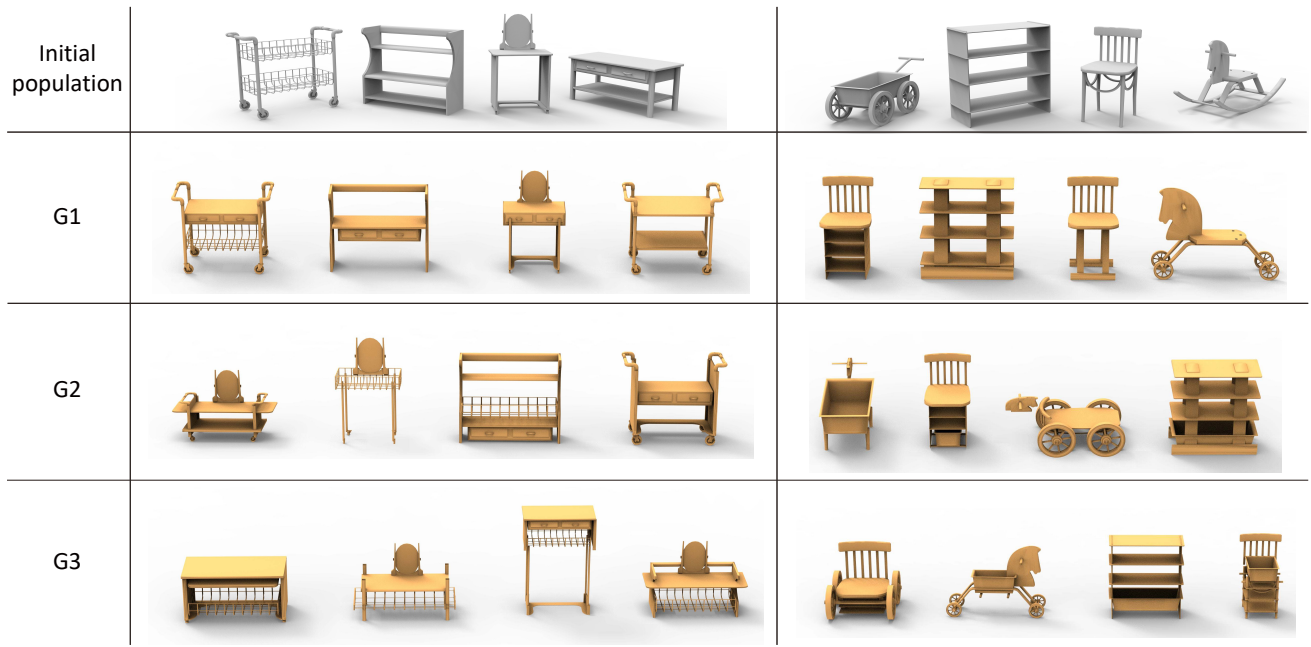}
    \caption{A gallery of modeling results from unconstrained evolution obtained with our functionality-aware approach. The first row shows two initial populations of four objects each, one per column. The next three rows show selected offspring from subsequent generations. The generations contain $78$, $113$, and $52$ shapes in total for the set in the first column, and $31$, $96$, and $45$ shapes for the set in the second column. 
    }
    \label{fig:unconstrained}
\end{figure*}

%% file: figures/constrained.tex
\begin{figure}[t]
    \centering
    \includegraphics[width=0.95\linewidth]{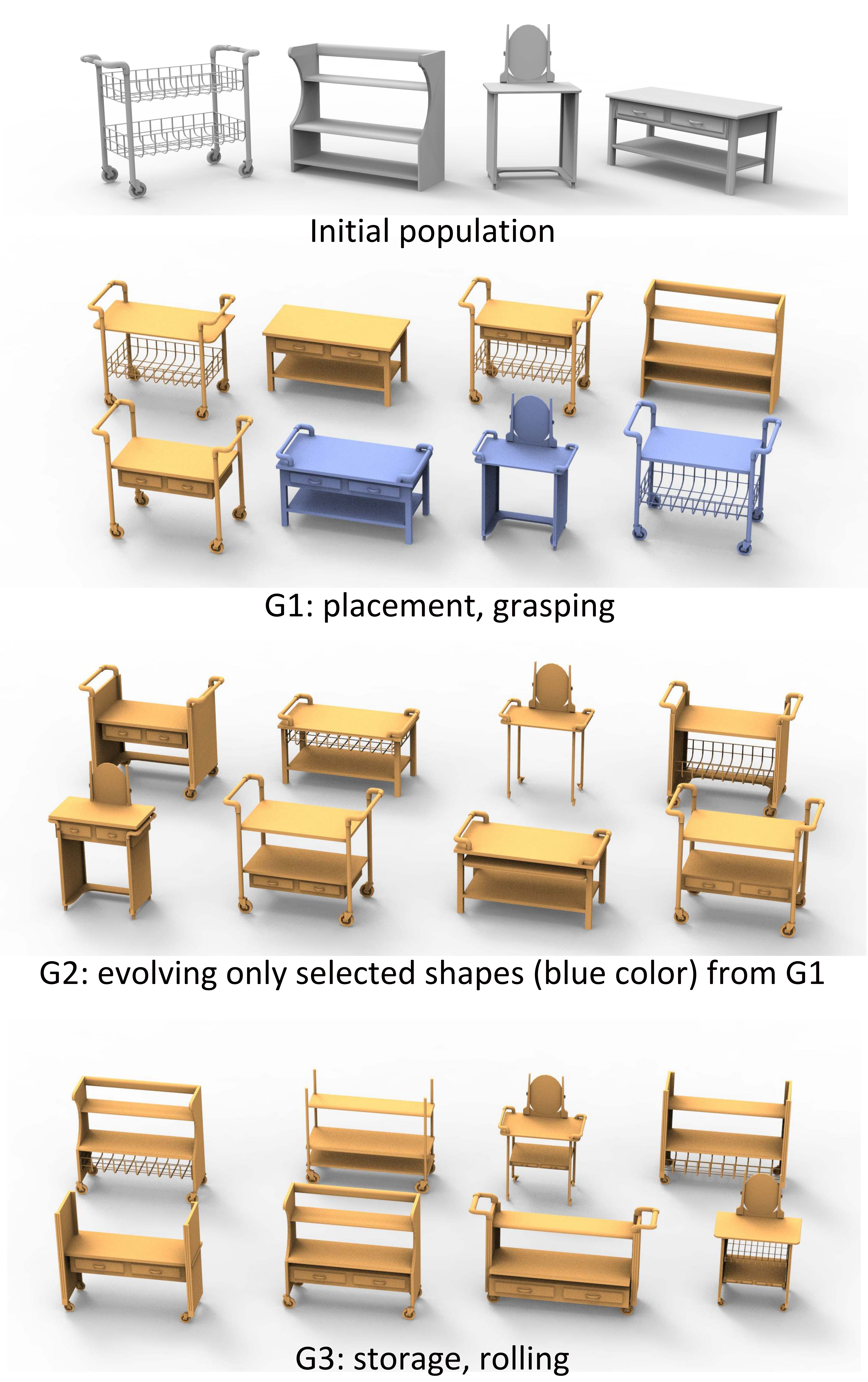}
    \caption{Results of constrained evolution by our functionality-aware modeling tool. The user evolves the initial population by constraining the offsprings with the functionality labels \emph{placement} and \emph{grasping}, obtaining the first generation G1. The user then selects the shapes marked in blue in G1 to be further evolved, to get G2. Finally, the functionality constraints \emph{storage} and \emph{rolling} are included as new preferences into the evolution of all the shapes, to obtain the third generation G3.}
    \label{fig:constrained}
\end{figure}

%% file: figures/multi-func.tex
\begin{figure}
    \centering
    \includegraphics[width=\linewidth]{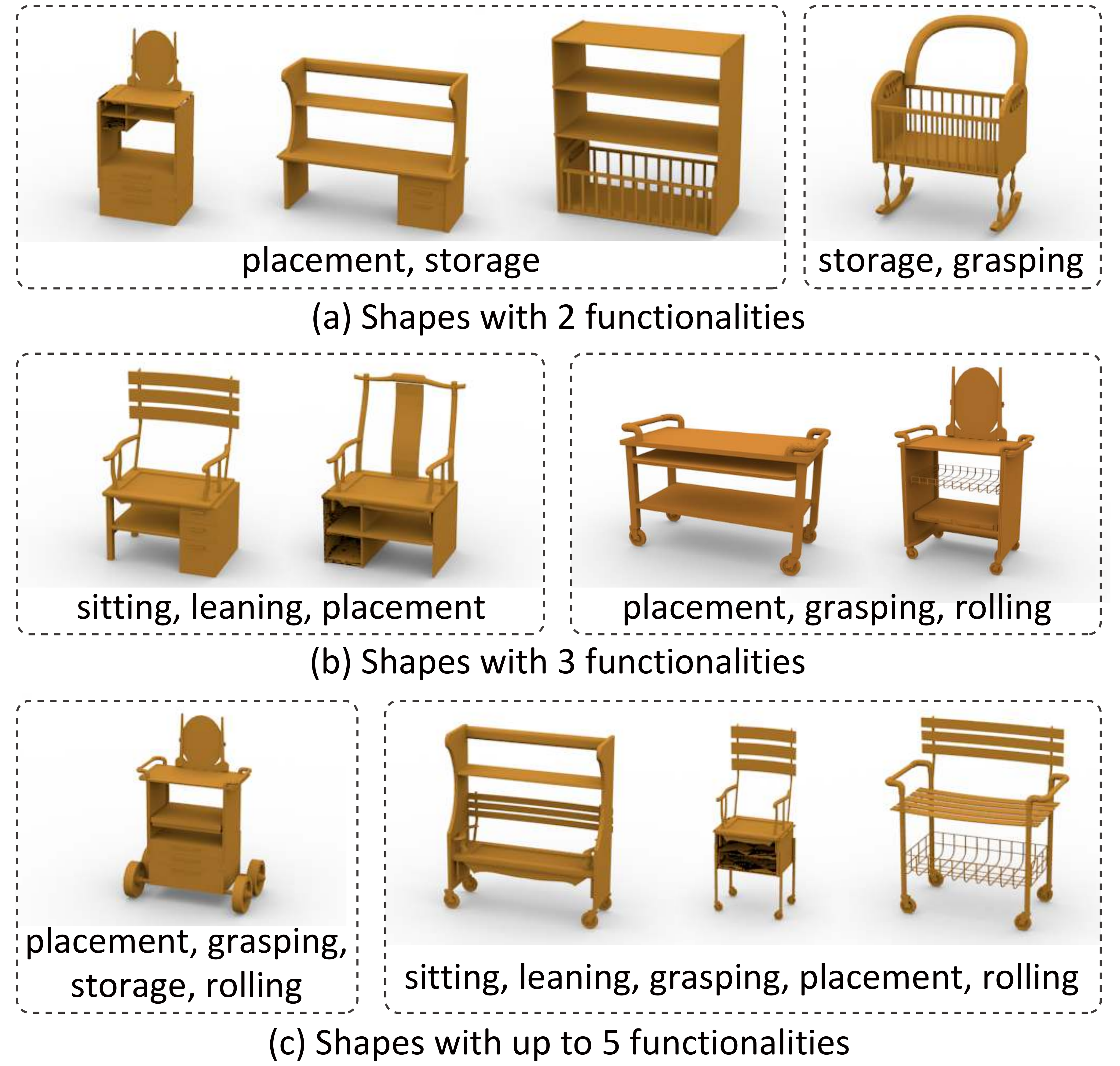}
    \caption{Results of model evolution where the objects are constrained to possess $2$, $3$, and up to $5$ functionalities.}
    \label{fig:multi-func}
\end{figure}

%% file: figures/large_set.tex
\begin{figure}[t]
    \centering
    \includegraphics[width=\linewidth]{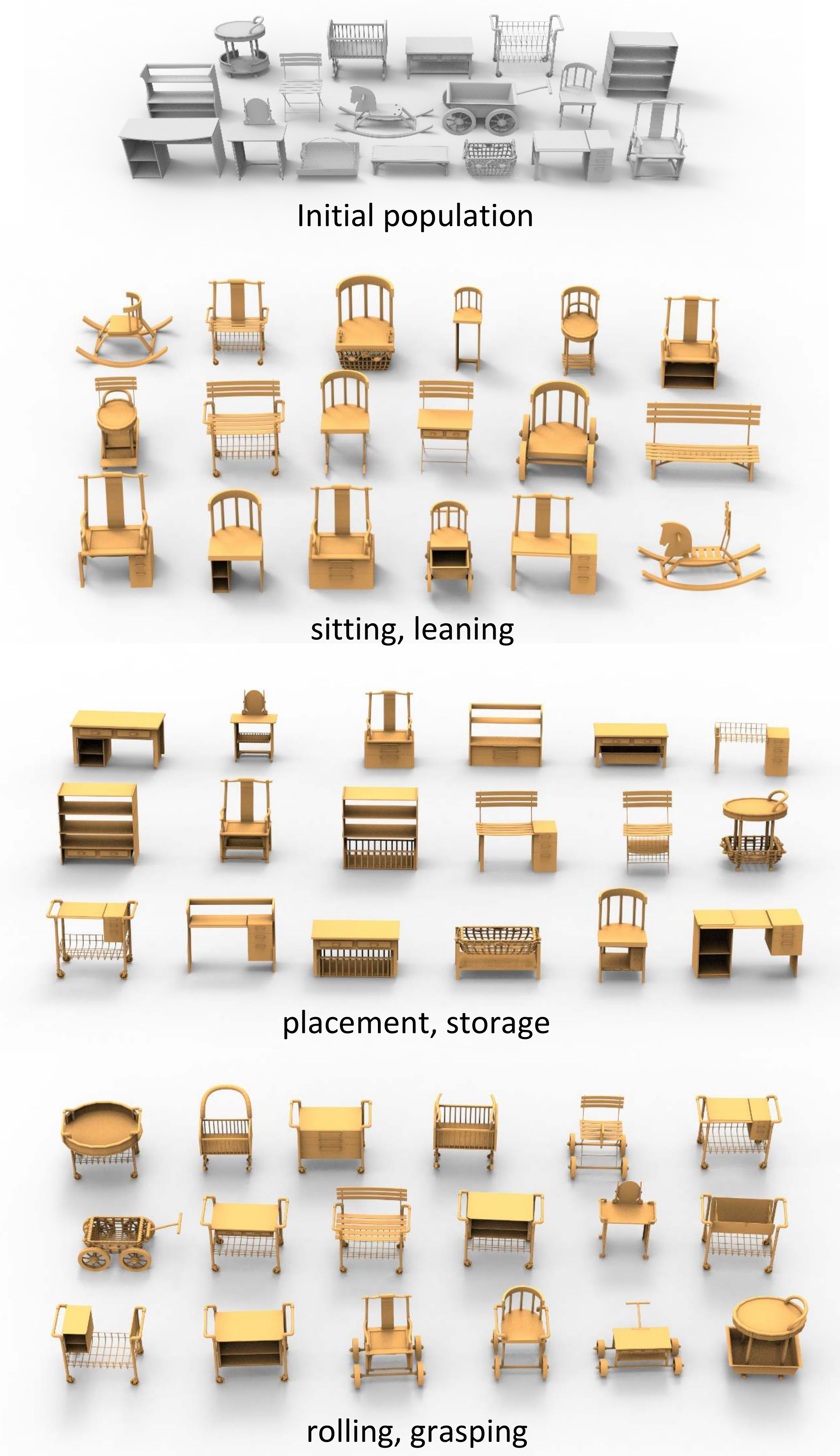}
    \caption{Results of model evolution from a large initial population ($17$ shapes) and with various functionality constraints to demonstrate scalability. The following populations are generated: \emph{sitting} + \emph{leaning} with $141$ shapes, \emph{placement} + \emph{storage} with $234$ shapes, and \emph{rolling} + \emph{grasping} with $258$ shapes. Only the top $18$ shapes for each set are shown, according to the ranking by functional plausibility.}
    \label{fig:large_set}
\end{figure}

%% file: figures/rank.tex
\begin{figure}[t]
    \centering
    \includegraphics[width=\linewidth]{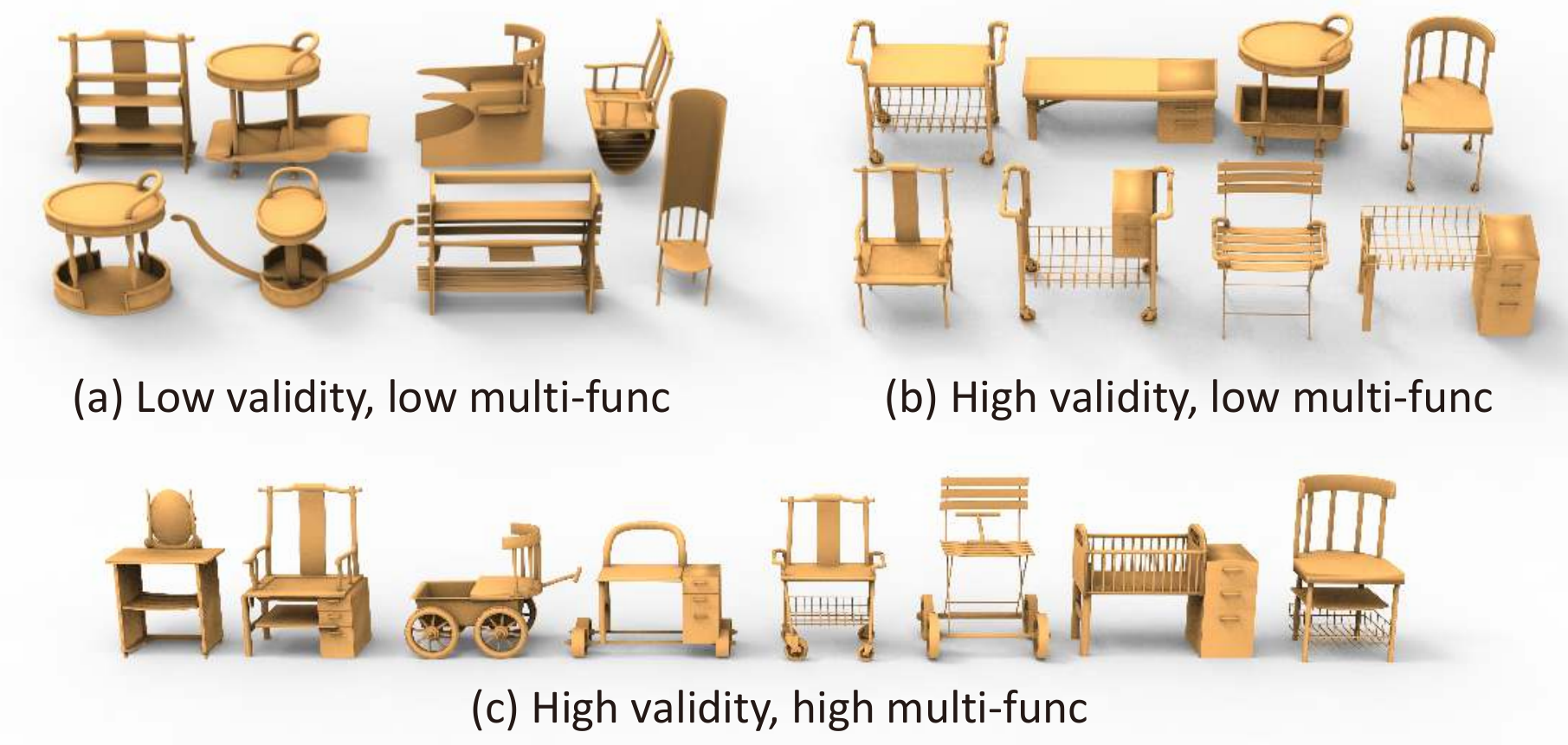}
    \caption{Comparison of ranking scores. Three sets of objects with different levels of priority for plausibility and multi-functionality measures.}
    \label{fig:rank}
\end{figure}

%% file: figures/zheng_comparison.tex
\begin{figure}[t]
    \centering
    \includegraphics[width=\linewidth]{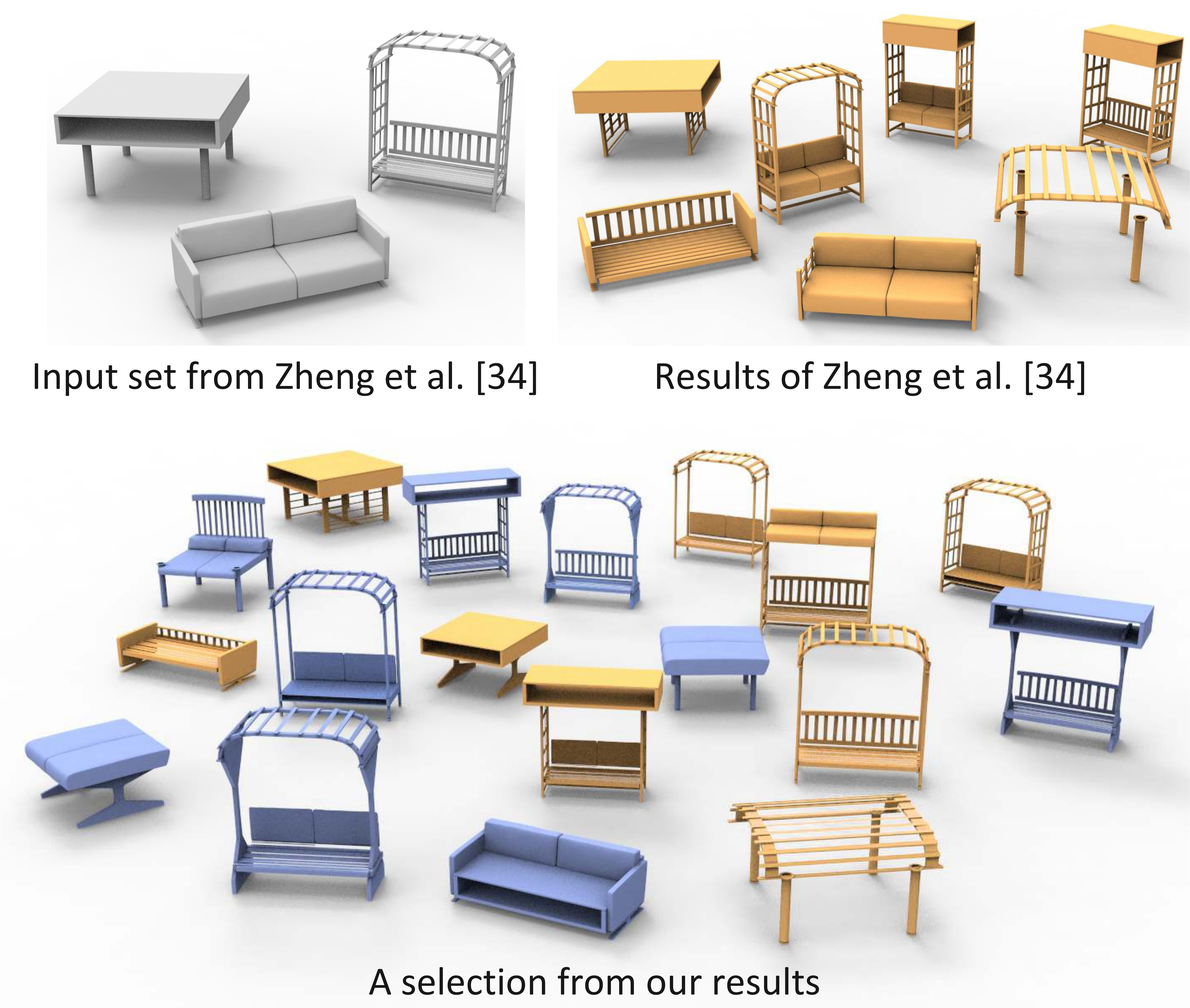}
    \caption{A comparison of our shape generation results to those from Zheng et al.~\cite{zheng13}, on an input set from their work. The set of offsprings generated by our method contains not only shapes producible by their method (shapes in yellow), but also other shapes (in blue) which their method cannot produce for various reasons discussed in the text.
    }
    \label{fig:zheng_comparison}
\end{figure}

%% file: figures/fu_comparison.tex
\begin{figure}[t]
    \centering
    \includegraphics[width=0.95\linewidth]{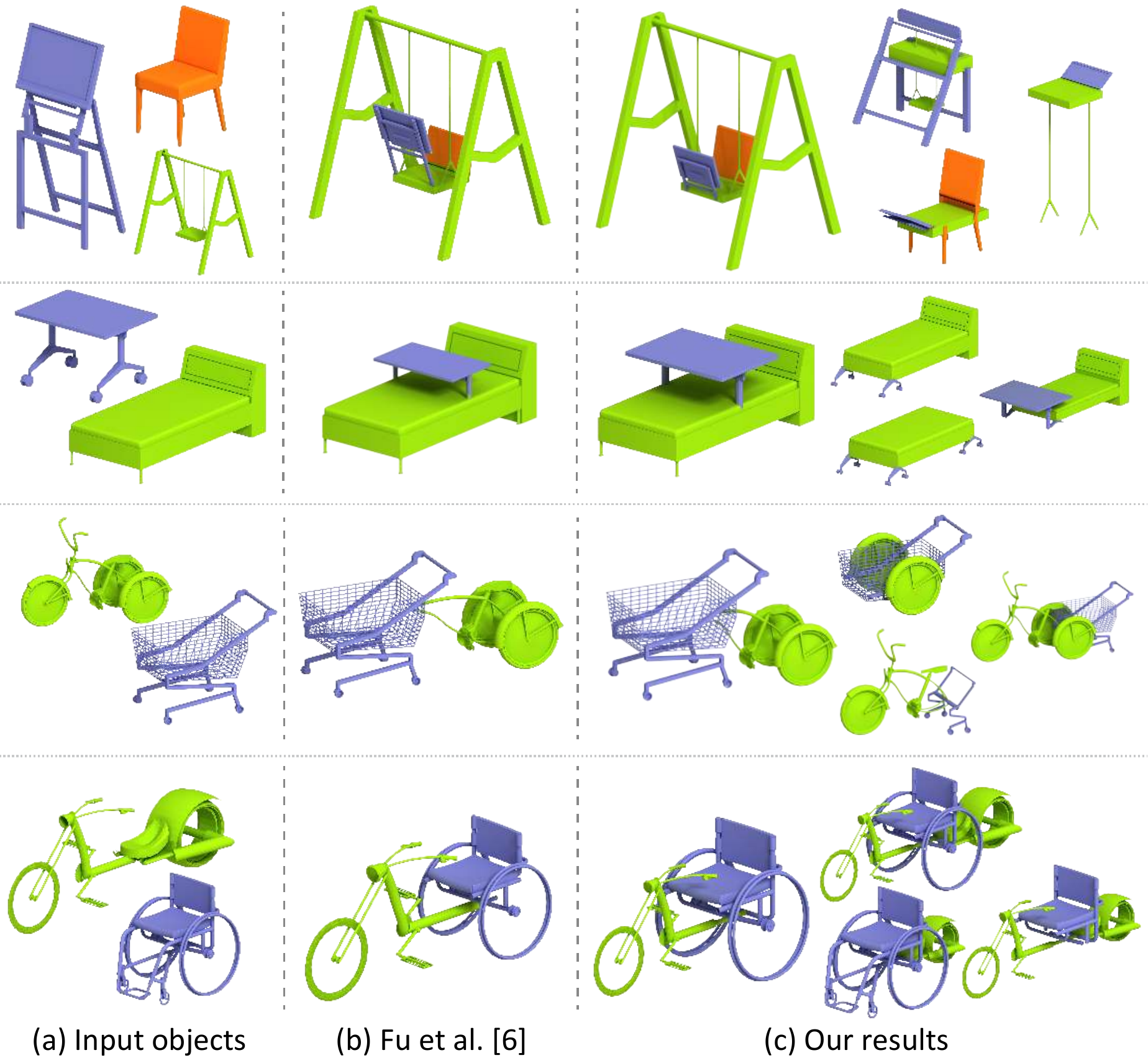}
    \caption{Comparison to functional hybrid generation by Fu et al.~\cite{fu2017}. In each row, we show the 3D shapes identified by their method (left) that match a human pose and the hybrid shape produced (middle). Using the same 3D shapes, our method is able to generate a more diverse set of hybrids (right), \emph{including} one that well resembles the outcome from their method, without a human pose as constraint.}
    \label{fig:compare_fu}
\end{figure}

%% file: figures/data_aug.tex
\begin{figure}[t]
    %
    \centering
    \includegraphics[width=0.9\linewidth,trim={10 10 10 10},clip]{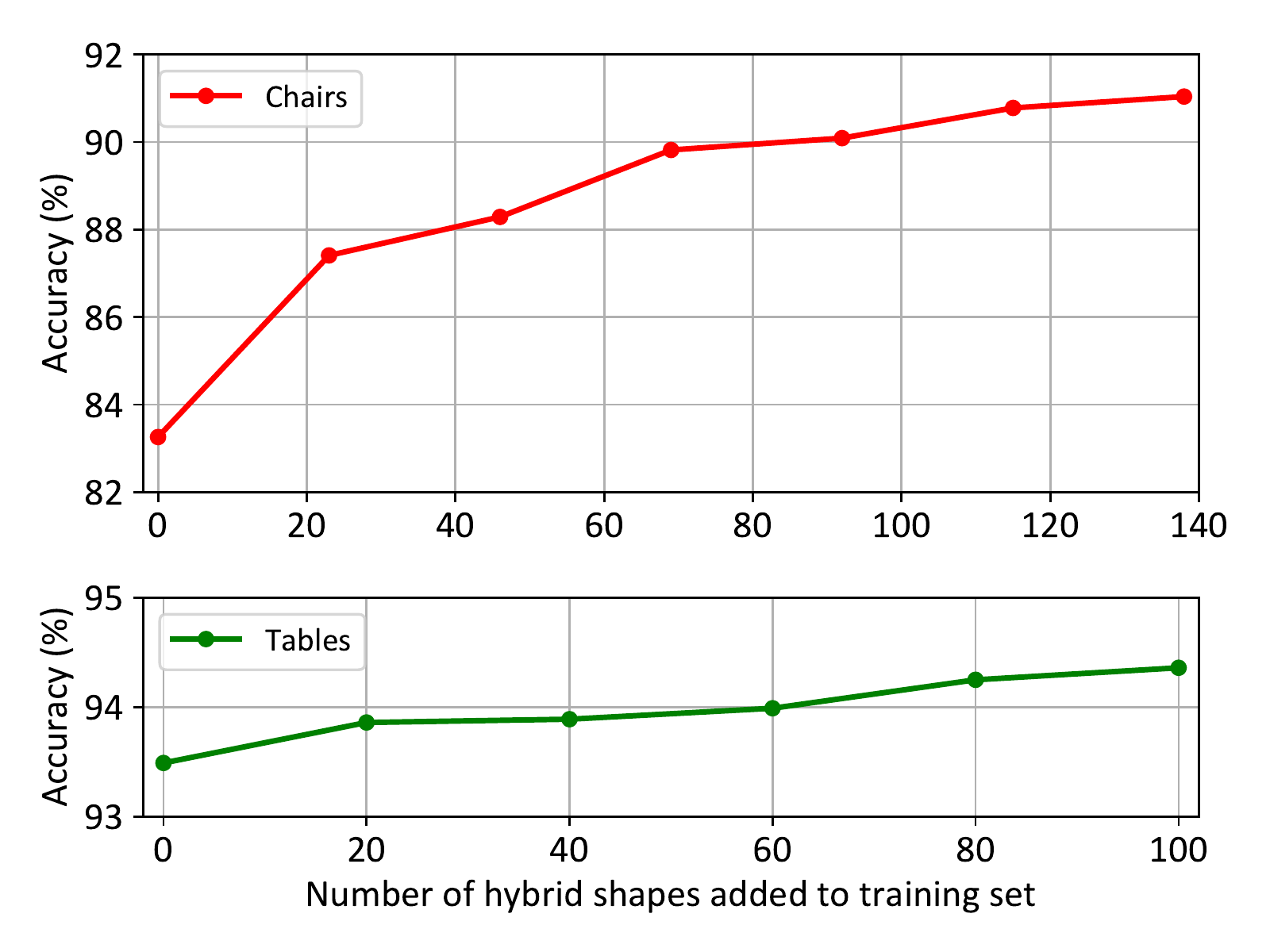}
    \caption{Results showing improved accuracy via data augmentation for learning shape segmentation, using PointNet, for two sets of shapes. The ShapeNet training set is augmented progressively with shapes evolved using our tool. Please refer to the text for details.}
    \label{fig:data_aug}
\end{figure}

%% file: figures/data_aug_visual.tex
\begin{figure}[t]
    \centering
    \includegraphics[width=0.95\linewidth,trim={30 30 30 60},clip]{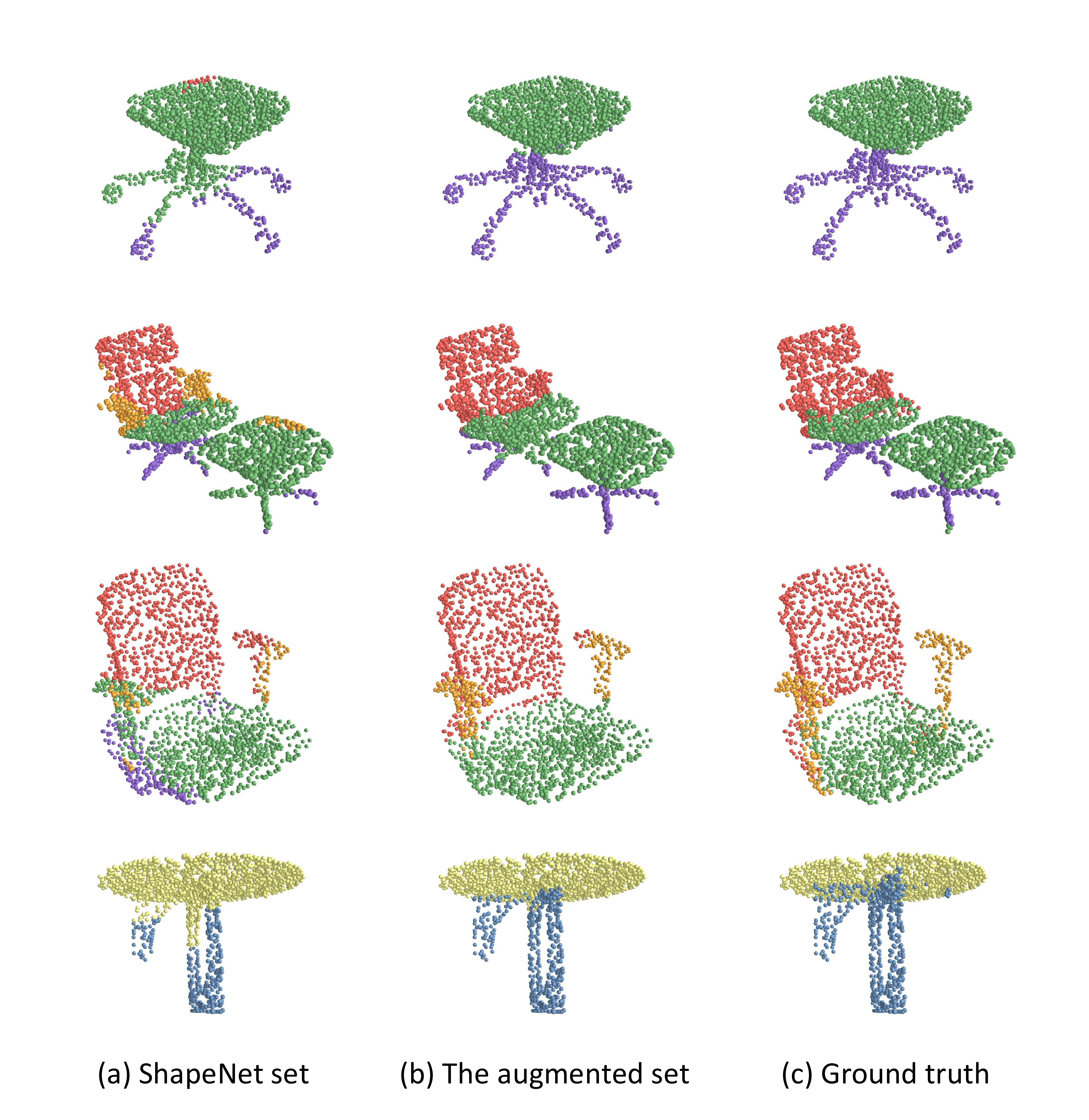}
    \caption{Visual results of PointNet segmentation on partial test shapes, using ShapeNet training vs.~the augmented training set (ShapeNet + our shapes).
    }
    \label{fig:data_aug_visual}
\end{figure}

%% file: future.tex
\section{Discussion, Limitations, and Future Work}
\label{sec:future}

We present the first \emph{functionality-aware}, \emph{user-in-the-loop} modeling tool to evolve a 
set of 3D objects, aimed at producing large and diverse sets of functionally plausible offsprings. 
Our work incorporates functionality analysis into the ``fit and diverse'' set evolution 
framework of Xu et al.~\cite{xu2012}. However, rather than restricting part exchange 
to objects within the same category as in Xu et al.~\cite{xu2012} and all the 
works so far on structure-preserving modeling~\cite{mitra2013}, our method
excels at generating \emph{cross-category} hybrids while allowing structure breaking. 
In the end, our evolutionary modeling tool makes a promising step towards 3D content
generation to achieve \emph{volume}, (intra-class) \emph{variety}, and (inter-class)
\emph{variation}, producing generations and generations of within- and cross-category hybrids via
controlled stochastic shape crossover.
%
%
But we re-iterate that our modeling tool only aims to suggest rough design prototypes, not refined 
final products. 
Fine-tuned modeling can be applied to selected, fully-functional offsprings in post-processing. 
%

\mypara{Hybridizing functionality models}
Since cross-category hybrids may belong to new functional categories beyond those available
in our knowledge base, it would be ideal to hybridize functionality models. In our current work,
we only hybridize at the object level, not at the functional category level. Instead of validating the
functionality of a new hybrid against a new functional category it rightfully belongs to, we only 
compare it partially to known functionality models since the hybrid is a crossover between objects 
covered by these models. To combine existing functionality models~\cite{hu2016} into new models
would require sufficient object data and their interaction contexts~\cite{hu2015}. Furthermore, the
resulting hybridized functionality models are generally not unique.

%


\mypara{Technical limitations}
%
There is still plenty of room to improve our current method from a technical perspective.
Our main technical limitations stem from our set goal of only producing rough design
prototypes, where the emphasis has been on functional properties of the offsprings,
not their precise geometries. Specifically, our evolution operator only produces
non-uniform part (group) scaling, while free-form deformations should allow a richer 
variety of offsprings. Moreover, our current part connection mechanisms still lack
an understanding of shape semantics and are unable to resolve topological mismatches or merge the meshes of parts together.
At the same time, our current functional plausibility score and constrained shape
evolution only offer a starting point for further investigation and development.
Last but not the least, we provide the option of computing plausibility scores with a detailed functionality partial matching. However, the computation relies on an expensive search procedure which is slow. Thus, directions for future work include more efficient manners of detecting partial functionalities of shapes, including the use of learning.


\mypara{Future work}
Aside from addressing the above limitations,
we would also like to explore other functionality modeling paradigms. For example,
instead of performing functionality-preserving style transfer, as in the work by Lun et
al.~\cite{lun2016}, we can invert the problem to style-preserving \emph{functionality 
transfer}: to transfer the extracted functionalities from a given shape to another shape, 
while preserving its stylistic features. Our evolution-based modeling offers a partial solution 
to this problem. Specifically, a crossover between two shapes is implicitly transferring 
functionalities associated with the exchanged part groups between the shapes.
Another interesting problem to study is \emph{functional analogy}, i.e., to synthesize a new 
shape ${\mathcal{S}_B}^\prime$ from an input $\mathcal{S}_B$ by adding functionalities which would separate a given 
shape ${\mathcal{S}_A}^\prime$ from a given shape $\mathcal{S}_A$. 
Finally, another direction for future work is to combine existing generative models based on neural networks with a form of functionality evaluation, to yield functionality-aware generative models of 3D shapes or scenes~\cite{khetarpal20,nagarajan20}.



%% file: ack.tex
\ifCLASSOPTIONcompsoc
  \section*{Acknowledgments}
\else
  \section*{Acknowledgment}
\fi

We thank the reviewers for their comments and suggestions. This work was supported in part by NSERC Canada (611370, 611649, 2015-05407), NSFC (61872250, 62032011), and a gift funding from Adobe.